\documentclass[apj,iop,numberedappendix]{emulateapjtweak}
\usepackage{apjfonts}

\usepackage{amsmath}
\usepackage{amsfonts}
\usepackage{amssymb}
\usepackage{natbib}
\usepackage{graphicx}
\usepackage{hyperref}
\hypersetup{pdfborder={0 0 0}, colorlinks=true, citecolor=blue, linkcolor=blue, pdftitle=Wave-Turbulence-Driven Coronal Heating}

\usepackage{booktabs}
\usepackage{cleveref}
\usepackage{relsize}
\usepackage{mathtools}

\usepackage{color}
\definecolor{midgreen}{RGB}{0,150,0}
\definecolor{darkgreen}{RGB}{0,100,0}
\definecolor{midred}{RGB}{215,0,0}
\definecolor{darkred}{RGB}{175,0,0}

\def\uh{$\text{ergs}\ \text{cm}^{-3}\ \text{s}^{-1}$}
\def\uf{$\text{ergs}\ \text{cm}^{-2}\ \text{s}^{-1}$}
\def\un{$\text{cm}^{-3}$}
\def\kms{$\text{km}\ \text{s}^{-1} $}

\newcommand{\neu}[3]{$#1\times\!10^{#2}\,$#3}

\creflabelformat{equation}{#2#1#3} % equation references don't come with brackets
\crefname{equation}{Eq.}{Eqs.}
\Crefname{equation}{Eq.}{Eqs.}
\crefname{section}{Section}{Sections}
\Crefname{section}{Section}{Sections}
\crefname{appendix}{Appendix}{Appendices}
\Crefname{appendix}{Appendix}{Appendices}
\crefname{table}{Table}{Tables}
\Crefname{table}{Table}{Tables}
\crefname{figure}{Figure}{Figures}
\Crefname{figure}{Figure}{Figures}

\makeatletter
\renewcommand\@makefntext[1]{\leftskip=2em\hskip-2em\@makefnmark#1}
\makeatother
\newlength{\cdonecol}
\newlength{\cdfullpage}
\newcommand{\zpm}{z_\pm}

\shorttitle{Wave-Turbulence-Driven Coronal Heating}

\shortauthors{Downs et al.}

\begin{document}

%--------------------------------------------------------------------------------------------------------------------------------------------
% Title/Header
%--------------------------------------------------------------------------------------------------------------------------------------------
\title{Closed-Field Coronal Heating Driven by Wave Turbulence}

\author{Cooper Downs\altaffilmark{1}, Roberto Lionello\altaffilmark{1}, Zoran Miki\'c\altaffilmark{1}, Jon A. Linker\altaffilmark{1}, Marco Velli\altaffilmark{2}}
\email{cdowns@predsci.com}
\altaffiltext{1}{Predictive Science Incorporated, 9990 Mesa Rim Rd. Suite 170, San Diego, CA 92121, USA}
\altaffiltext{2}{EPSS, UCLA, Los Angeles, CA 90095, USA}

%\slugcomment{Draft \today}
\slugcomment{Accepted to ApJ, September 30, 2016}

%--------------------------------------------------------------------------------------------------------------------------------------------
% Abstract
%--------------------------------------------------------------------------------------------------------------------------------------------
\begin{abstract}
\label{section:abstract}
To simulate the energy balance of coronal plasmas on macroscopic scales, we often require the specification of the coronal heating mechanism in some functional form. To go beyond empirical formulations and to build a more physically motivated heating function, we investigate the wave-turbulence-driven (WTD) phenomenology for the heating of closed coronal loops. Our implementation is designed to capture the large-scale propagation, reflection, and dissipation of wave turbulence along a loop. The parameter space of this model is explored by solving the coupled WTD and hydrodynamic evolution in 1D for an idealized loop. The relevance to a range of solar conditions is also established by computing solutions for over one hundred loops extracted from a realistic 3D coronal field.  Due to the implicit dependence of the WTD heating model on loop geometry and plasma properties along the loop and at the footpoints, we find that this model can significantly reduce the number of free parameters when compared to traditional empirical heating models, and still robustly describe a broad range of quiet-sun and active region conditions. The importance of the self-reflection term in producing relatively short heating scale heights and thermal nonequilibrium cycles is also discussed.
\end{abstract}


%--------------------------------------------------------------------------------------------------------------------------------------------
% Introduction
%--------------------------------------------------------------------------------------------------------------------------------------------
\section{Introduction}
\label{section:introduction}

Understanding the mechanism by which the solar corona is heated to millions of degrees is a long-standing problem in solar astrophysics. The physical mechanisms have been hotly debated for decades, and pinning them down is of great scientific importance. Identifying a heating mechanism that can be implemented into large-scale simulations will also advance the development of realistic, physics-based models of the solar corona.

Numerous theories have been developed for this problem. Two modern theoretical perspectives are given in the introduction of \citet{van_ballegooijen14} and the review by \citet{klimchuk15}. In short, coronal heating theories are typically classified in two main groups; mechanism involving wave dissipation, known as `alternating current' (AC) theories, and those involving the buildup of magnetic stresses, known as `direct-current' (DC) theories. Because of the immense complexity of realistic coronal plasmas, progress for a given theory typically evolves from analytic theory \citep{hollweg78_waves,parker72}, to sophisticated modeling for idealized configurations \citep{rappazzo08,van_ballegooijen11,pontin15}, to implementations suitable for 3D magnetohydrodynamic (MHD) models \citep{vanderholst14,bingert11}. All the while our observational capabilities have improved, and work has strived to provide constraints to proposed mechanisms \citep{mandrini00,warren11,schmelz15}. 

A complementary approach to this problem has been the development of empirical heating models or scaling laws that can be used in hydrodynamic (HD) or MHD models. In this case an empirically formulated heating rate based on local plasma or magnetic loop properties can be used to study the thermodynamic response of plasma to heating \citep[e.g.][]{antiochos99_loop,aschwanden02,muller05,mikic13}, and to compare model results directly to observations \citep[][]{schrijver04,lundquist08,lionello09,downs10,mok16}. This approach can also be used for more general applications, where a realistic thermodynamic state in the corona is required to model other coronal phenomena \citep[][]{downs12,downs13,jin13}.

Observationally there is strong evidence that turbulent Alf\'enic fluctuations are present and possibly dissipated in the low solar corona \citep{tomczyk09,bemporad12,hahn12,hahn14}. There are also indicators that hydrodynamic evolution along coronal loops can be quite common, especially for active regions. These include observations of coronal rain and catastrophic cooling cycles \citep{schrijver01,antolin12,antolin15_multistranded_multitemp_rain}, as well as pervasive cooling signatures observed in extreme ultraviolet  \citep{viall11,viall12,auchere14,froment15} and soft X-Ray imaging \citep{ugarte-urra06}. A successful heating model should be plausibly consistent with such observations.

In a perfect world we would model coronal dynamics with a fully self-consistent theory, but in practice this is not yet possible. On the other hand, it is preferable to employ physics-based models of coronal heating that can be used for broader modeling applications. In this spirit, we investigate a wave-turbulence-driven (WTD) phenomenology based on the work of \citet{velli93} and \citet{verdini10} that uses the propagation, reflection, and dissipation of Alf\'enic turbulence to heat the corona. We couple the WTD phenomenology to a 1D model of the time-dependent hydrodynamic evolution of plasma along a closed coronal loop. Our approach is designed to investigate the basic physical behavior of the model, and examine its scaling in the context of key aspects of coronal heating, such as the heating power, scale height, apex temperatures, and ensuing dynamics. Our analysis also investigates the performance of the model for a range of quiet and active coronal conditions and compares it to empirical scaling laws. 

The paper is organized as follows: Section~\ref{section:model} describes the coupled HD/wave system that comprise the WTD model and our numerical approach. In Section~\ref{section:analytic} we illustrate the behavior of the relevant terms in the wave system, and Section~\ref{section:heating_analytic} presents an analysis of the salient properties relevant to coronal heating. We provide example solutions to the full system of equations in Section~\ref{section:single_loop}, and Section~\ref{section:paramspace} is used to explore the scaling of the model on the key loop parameters and the free parameters of the model. In Section~\ref{section:realistic_loops} we apply the model to a broad selection of loops traced from a realistic 3D magnetic field and assess the scaling of the model. We conclude in Section~\ref{section:conclusion}.

%--------------------------------------------------------------------------------------------------------------------------------------------
% Model
%--------------------------------------------------------------------------------------------------------------------------------------------
\section{Model Description}
\label{section:model}

In order to explore the viability of a WTD-based heating model for arbitrary configurations of closed coronal loops we employ the MHD1\_LOOP code. This code was first used by \citet{lionello02} and described in detail recently by \citet{mikic13}.  MHD1\_LOOP solves the following 1D mass, momentum, and energy and equations for plasma hydrodynamics,
\allowdisplaybreaks
\begin{align}
%-------------------------
% Hydro Equations
% Mass
\label{eq:hydro:mass}
&\frac{\partial\rho}{\partial{t}}+\frac{1}{A}\frac{\partial}{\partial{s}}\left ( A\rho u \right )=0,\\
% Momentum
\label{eq:hydro:momentum}
&\rho \left ( \frac{\partial u}{\partial{t}}+u\frac{\partial u}{\partial{s}}\right )= -\frac{\partial p}{\partial{s}}-\rho g(s)+\frac{1}{A}\frac{\partial}{\partial{s}}\left (A\rho\nu\frac{\partial u}{\partial{s}}\right ),\\
% Energy
\label{eq:hydro:energy}
\begin{split}
&\frac{\partial T}{\partial{t}}+\frac{1}{A}\frac{\partial}{\partial{s}}\left ( ATu \right )=-(\gamma-2)\frac{T}{A}\frac{\partial}{\partial{s}}\left ( Au \right )\\
&\quad +\frac{ (\gamma-1)}{2k_Bn_e}\biggl[ \frac{1}{A}\frac{\partial}{\partial{s}}\left ( A\kappa_\|\frac{\partial T}{\partial{s}}\right )-n_e^2\Lambda(T)+Q_{w}\biggr],\\
\end{split}
\end{align}
along the loop coordinate $s$, which ranges from 0 to the loop length, $L$, and $A$ is the loop area, $A(s)=A_0B_0/B(s)$, where $A_0$ is the physical flux tube area at $s=0$ and $B_0$ is the corresponding magnetic field there.  Arbitrary forms of loop geometry and areal expansion can be incorporated by tracing field lines from a 2D or 3D magnetic vector field. The local orientation of the field provides the component of gravity parallel to the loop, $g(s)=\vec{g}\cdot\hat{b}$, and the variation of the local field magnitude provides the areal expansion factor, $A/A_0$. By convention, in the rest of the paper we refer to the $s\!=\!0$ and $s\!=\!L$ boundaries as the left and right footpoints respectively. We orient the solutions as going from positive $B_r$ to negative $B_r$, where $B_r$ is the component of the magnetic field with respect to the solar surface.

In the 1D momentum equation (\cref{eq:hydro:momentum}) the flow velocity, $u$, is always along the field and becomes a scalar quantity. The kinematic viscosity term, $\nu$, is used to damp out unresolved hydrodynamic waves and is relatively small, corresponding to a diffusion time of $L^2/\nu\sim 2000$ hr.  For simplicity we model a single-fluid hydrogen plasma with an adiabatic index of $\gamma=5/3$. The mass density relation becomes $\rho=m_pn_e$, where $m_p$ is the proton mass and $n_e$ is the election number density. The pressure relation is $p=2k_Bn_eT$, where $T\equiv T_e$ is the electron temperature (assumed equal to the proton temperature), and $k_B$ is the Boltzmann constant.

The second row of terms in the energy equation (\cref{eq:hydro:energy}) specify the non-ideal terms relevant to coronal energy balance, which are parallel electron heat conduction, optically thin radiative losses, and the coronal heating term. A thorough discussion of the conduction and loss terms is given in \citet{mikic13}, which we only summarize here. First, the classical form of Spitzer thermal conductivity, $\kappa_\|(T)=\kappa_0T^{5/2}$, with $\kappa_0= 9\times \text{10}^{-7}$ is used for coronal temperatures above a cutoff value, $T_c=$350,000 K. For $T<T_c$, $\kappa_\|(T)$ and the radiative loss rate, $\Lambda(T)$, are modified such that product of the two remains unchanged. This modification serves to significantly broaden the transition region scale lengths from several kilometers to a hundred or so kilometers, while leaving the coronal solution relatively unchanged \citep[see also][and discussions therein]{abbett07,lionello09,downs10}. While not strictly necessary for 1D solutions, where using large numbers of gridpoints or adaptive mesh refinement methods is possible \citep[such as][]{antiochos99_loop}, this broadening method allows for a modest number of gridpoints to be used per run (1000-2000). This reduces simulation run times to the order of minutes per loop, allowing for rapid parameter-space scanning. This technique is also directly relevant to 3D thermodynamic HD and MHD simulations of the corona where resolving the unbroadened scales is not yet feasible \citep[e.g.][]{mok08,lionello09,downs10,vanderholst14}.

The optically thin radiative loss rate, $\Lambda(T)$, can either be specified as a parametrized curve \citep[e.g.][]{athay86}, or calculated from synthetic spectra. In this study, we use $\Lambda(T)$ calculated from the CHIANTI 7.1.3 atomic database \citep{dere97,landi13_chianti} using `hybrid' coronal abundances \citep{schmelz12}, and the CHIANTI ionization equilibrium model \citep{dere09}. Similar to \citet{mikic13}, we smoothly reduce $\Lambda (T)$ to zero as $T$ approaches low chromospheric temperatures. In this case the smooth reduction spans from $T=$\neu{1.0\!-\!4.0}{4}{K}. Instead of forcing $\Lambda (T)$ to be identically zero at the boundary temperature ($T_0=$\neu{1.75}{4}{K}) we add the chromospheric heating term described by \citet{sokolov13}. This exactly balances the radiative losses at $T_0$ with an exponential heating term defined by the short chromospheric scale height, $\lambda (T_0) = 1.1$ Mm. It allows the solution to find a simple hydrostatic equilibrium in the model chromosphere and decays long before it is energetically relevant to the corona.

\subsection{WTD Heating Model}
\label{section:wtd_specification}

The remaining term in the energy equation is the primary heating term, $Q_w$, which is determined by choosing a model for the time-dependent propagation and dissipation of wave turbulence. A primary goal of this study is to explore the potential unification of a wave-turbulence model that describes the heating and acceleration of the fast solar wind with one that describes heating in the low corona. To this end, we employ the same formulation for the evolution of low-frequency Alfv\'en wave turbulence as \citet[][a solar wind study]{lionello14}, which is based on the work of \citet{velli93} and most recently \citet{verdini10}. Invoking symmetry in the perpendicular direction to the mean field and the limit of low frequencies ($\omega \to 0$), the evolutionary equations can be  expressed in terms of the scalar magnitude of the Els\"asser variables, $z_\pm = \delta u \mp \delta b / \sqrt{4\pi\rho}$, and take the following form:
\begin{equation}
\label{eq:zpzm}
%-------------------------
% zpzm Equations
\frac{\partial z_\pm}{\partial{t}}+(u\pm v_A)\frac{\partial z_\pm}{\partial{s}} =  R_1 z_\pm + R_2 z_\mp - \frac{z_\pm |z_\mp|}{2\lambda_\perp},
\end{equation}
with the following definitions,
\\
\begin{align}
R_1 &= \frac{1}{4}(u\mp v_A)\frac{\partial \ln{\rho}}{\partial{s}},\\
R_2 &= \frac{1}{2}(u\mp v_A)\frac{\partial \ln{v_A}}{\partial{s}},\\
\label{eq:heat_def}
Q_{w} &= \rho \frac{|z_-|z_+^2+|z_+|z_-^2}{4\lambda_\perp},\\
\label{eq:ew}
e_\pm &= \rho\frac{z_\pm^2}{4},\\
\label{eq:lambda}
\lambda_\perp &= \lambda_0 \sqrt{\frac{B_{W}}{B}},
\end{align}
where the $+$ and $-$ notations indicate fluctuations propagating parallel and anti-parallel to the magnetic field line respectively. The advective derivative plus the diagonal $R_1$ term describe the linear propagation of a given wave species, which alone would be equivalent to standard Wentzel-Kramers-Brillouin (WKB) approximation for Alfv\'en wave propagation \citep{jacques77}. The in situ generation of counter-propagating species enters through the off-diagonal reflection term, $R_2$, which becomes active in the presence of large-scale gradients in the Alfv\'en speed, $v_A=B/\sqrt{4\pi\rho}$. The phenomenological dissipation term, $ {z_\pm |z_\mp|}/{2\lambda_\perp}$ \citep{matthaeus99,dmitruk01}, activates in the presence of both species and specifies the nonlinear conversion of turbulent fluctuations into thermal energy. $e_\pm$ defines the wave energy of a given species, and multiplying \cref{eq:zpzm} by $\rho z_\pm/2$ and summing over both species leads to the total energy density dissipated per unit time, which specifies the heating term, $Q_w$, in \cref{eq:heat_def}. 

$\lambda_\perp$ is a parameter which describes the transverse correlation length of the fluctuations. $\lambda_\perp$ influences the effective timescale of dissipation, and evolutionary equations for it have been proposed with various degrees of complexity \citep[see][and references therein]{zank12}. For simplicity and consistency with previous studies, we chose to only follow the variation of $\lambda_\perp$ with the expansion factor (\cref{eq:lambda}), which typically dominates the evolution. To make the equation amenable to 3D MHD models, we choose a constant reference magnetic field, $B_W=6.09$ Gauss. It is equivalent to assuming that the correlation length at the base of the corona is not uniform everywhere but depends weakly on the base magnetic field strength. 

For an open flux tube, such as those studied by \citep{verdini10,lionello14}, heating strictly arises from the self-reflection of the outward propagating waves, while for a closed flux tube, interactions may arise both from self-reflection and via the counter-propagating waves launched at the opposite loop footpoint. In the latter case, the degree to which the fluctuations are correlated will determine their level of interaction and hence the dissipation rate. For simplicity we consider the case where the fluctuations are completely correlated, which allows us to track two wave species in total instead of four. The ramification of this choice is discussed briefly in Section~\ref{section:analytic_both_sides}.

In this theory, the dissipation term has a constant multiplier, of order unity, due to the unknown absolute timescales involved. This constant can be absorbed into $\lambda_0$, which we typically choose to range from $0.01-0.07\ \text{R}_s$ \citep[as in][]{verdini10,lionello14}. This range is slightly larger than recently reported correlation lengths determined observationally in the photosphere by \citet{abramenko13}, but it should not be taken too literally because of the simplicity of the model. For example,  \citet{van_ballegooijen11} found the true dissipation rate in a reduced MHD simulation of an idealized loop to be around a factor of four to five times smaller than the phenomenological dissipation rate. To match this dissipation rate, our choice of $\lambda_0$ in our model would have to be larger by the same factor.

Lastly, our WTD model is similar in spirit to the 3D MHD model described by \citet[][]{vanderholst14}, AWSoM. Both models use an evolutionary equation that includes the reflection and dissipation of Alfv\'enic fluctuations, and \cref{eq:zpzm} resembles an amplitude form of the AWSoM implementation. However, there is a key difference in the treatment of the linear reflection term: in AWSoM, reflection is attenuated in various ways that limit its role in the closed corona, while we do not limit it here. As we will show, this term plays an important role in determining the stratification of the WTD heating rate. This difference can lead to fundamentally different solutions in the two models.  In particular, it can dramatically affect the likelihood for the appearance of nonequilibrium solutions.

\subsection{Solution Scheme and Boundary Conditions}

The solution scheme and boundary conditions for the hydrodynamic equations (Eqs.~\ref{eq:hydro:mass}-\ref{eq:hydro:energy}) are relatively straightforward and the approach is the same as described in \citet{mikic13}. The code employs a non-uniform mesh spacing to capture fine gradients near and around the transition region and employs an operator split semi-implicit approach to advance the code at the advective time step. Each iteration amounts to a mix of explicit upwind advection advances and implicit tridiagonal solves.  

The parallel velocity at the boundary is specified using the method of characteristics, and the footpoint temperature is set a chromospheric value of $T_0=$\neu{1.75}{4}{K}. Our boundary number density is set to  $n_{e,0}=\ $\neu{6}{12}{\un}. This relatively large value is used to maintain a sufficient density reservoir in order to prevent total chromospheric evaporation for cases where large temperatures and/or dynamic evolution is involved.

For the wave amplitudes, we formulate our solution scheme by noting that the fast dynamical timescale implied by the wave propagation speeds (set by the Alfv\'en speed) is typically much shorter than the timescales of loop hydrodynamics (set by the flow and sound speeds). Therefore, for every semi-implicit iteration of the hydrodynamic equations at the advective CFL limit, we advance the wave amplitudes (\cref{eq:zpzm}) by sub-cycling each species together at the explicit limit implied by the wave speeds, $v_{w\pm}=u\pm v_A$, until we reach the new hydrodynamic time level. The advective term on the left hand side of \cref{eq:zpzm} is advanced using a 2\textsuperscript{nd} order flux-limiter scheme, and we use the symmetric OSPRE flux limiter function \citep{waterson2007_OSPRE}. The terms on the right hand side, including the coupled reflection and dissipation terms are solved using a point implicit method, which is connected to the explicit update via Strang splitting \citep[see][]{toth12}. The time update for each complete sub-cycle follows a 2nd order TVD Runge-Kutta scheme \citep{gottlieb1998_TVD_RungeKutta}.

The wave amplitude boundary conditions are set using simple pass-through conditions, where the waves coming into the simulation domain are set by a fixed amplitude $z_0$, and the outgoing waves propagate through without reflection at the boundary. Operationally this is done by setting $z_\pm=z_0$ when the wave speed of a species, $v_{w\pm}=u\pm v_A$, is positive at the left boundary (since $\vec{B}\!\cdot\!\vec{s}\!>\!0$) and negative at the right boundary ($\vec{B}\!\cdot\!\vec{s}\!<\!0$). When the wave speed for a species is oppositely directed (leaving the domain), $z_\pm$ is linearly extrapolated to the last half mesh point outside the boundary.

\subsubsection{Treating the Photosphere--Corona Transition}
\label{section:effective_rho}

Our goal is to develop a closed-field heating specification for application in global coronal models.  This precludes a detailed model of the physics in the photosphere and chromosphere.  The chromosphere is known to be highly inhomogeneous and dynamic \citep[e.g.][]{hansteen06,hansteen10,depontieu15,carlsson15}, with complex propagation and reflection of waves, and steepening of waves into shocks.  In addition to this complex behavior, it is in this layer that the flux tubes that eventually penetrate the corona expand strongly in area.  This expansion arises from the concentration of the magnetic field in the photosphere into individual strong flux elements.  With rising height, the plasma beta drops sharply, and these flux elements expand rapidly to fill the corona.  This expansion affects the propagation and reflection of Alfv\'en waves via the gradients of the Alfv\'en speed. For example, \citet{van_ballegooijen11,van_ballegooijen14} investigated in detail the effect of this expansion on coronal heating by Alfv\'en waves.

Rather than attempting to model the complex propagation and reflection of waves from the photosphere through the chromosphere and transition region, we choose to specify the wave amplitude {\it in the upper chromosphere}, at a height that is conveniently specified in terms of the value of the local plasma density.  In this paper we choose a density of \mbox{$n_{e}^{\text{eff}}\!=\,$\neu{2}{11}{\un}}.  Since we are required to use a significantly larger plasma density at the lower boundary of our domain (\mbox{$n_{e,0}\!=\,$\neu{6}{12}{\un}} in this work), and need to apply the boundary conditions on the wave amplitudes here, we choose to propagate the waves ideally, without reflection or dissipation, between these two layers.  We do this by smoothly setting the non-WKB terms in the propagation equations to zero for densities above $n_{e}^{\text{eff}}$, causing the waves to propagate according to WKB theory here. In this way, any wave energy specified at the boundary is simply propagated with no losses up to the effective height implied by $n_e^{\rm eff}$, regardless of the Alfv\'en speed profile in this region.  After going below the density threshold, the waves can be reflected and dissipated, and behave as if they were injected directly at this height.  

Of course, this approximation implies that we will not get any heating in the layers below $n_{e}^{\text{eff}}$, but, as mentioned, we are not attempting to model the chromosphere accurately. Besides, our approximation of lowering the radiative losses at chromospheric temperatures makes the model insensitive to the details of chromospheric heating.  Ultimately, this approach implies that the actual boundary condition for density $n_{e,0}$ has no influence on the wave energy flux (provided $z_0$ is scaled according to the WKB solution for $u\ll v_A$), and can be chosen by other considerations. With this simplified treatment, it is important to keep in mind that in our model, the observational constraint on the magnitude of the wave amplitudes is in the upper chromosphere, corresponding to $n_e^{\rm eff}$.

In the same context, we also explored the effect of different boundary conditions (BCs) for the wave amplitudes $z_+$ and $z_-$ \citep{verdini12}.  We tested ``pass-through'' BCs, described in the previous section, and also ``reflection'' BCs, for which waves are fully reflected at the boundary.  As mentioned above, the complicated propagation and reflection of waves in the region between the photosphere and upper chromosphere implies that the correct BCs for our problem are not uniquely known {\it a priori}; we need to select the boundary conditions that lead to the most robust injection of waves into the corona.  By extensive exploration we found that the pass-through BCs produced the most physical coronal heating solutions, and led to the most predictable scaling behavior.

%--------------------------------------------------------------------------------------------------------------------------------------------
% Analytic ZpZm
%--------------------------------------------------------------------------------------------------------------------------------------------
\section{Analysis of the $\protect{\zpm}$ Equation}
\label{section:analytic}

Before proceeding to numerical solutions of the coupled hydrodynamic and wave system, it is instructive to discuss the analytic and numerical behavior of the wave amplitude equation (\cref{eq:zpzm}) for a stationary background loop. This allows us to examine the contributions of each term to the $z_\pm$ solutions, and to illustrate what factors influence the net flux of energy entering into the domain through the loop footpoints.

\begin{figure*}[hbtp]
\centering
\includegraphics[width=\cdfullpage]{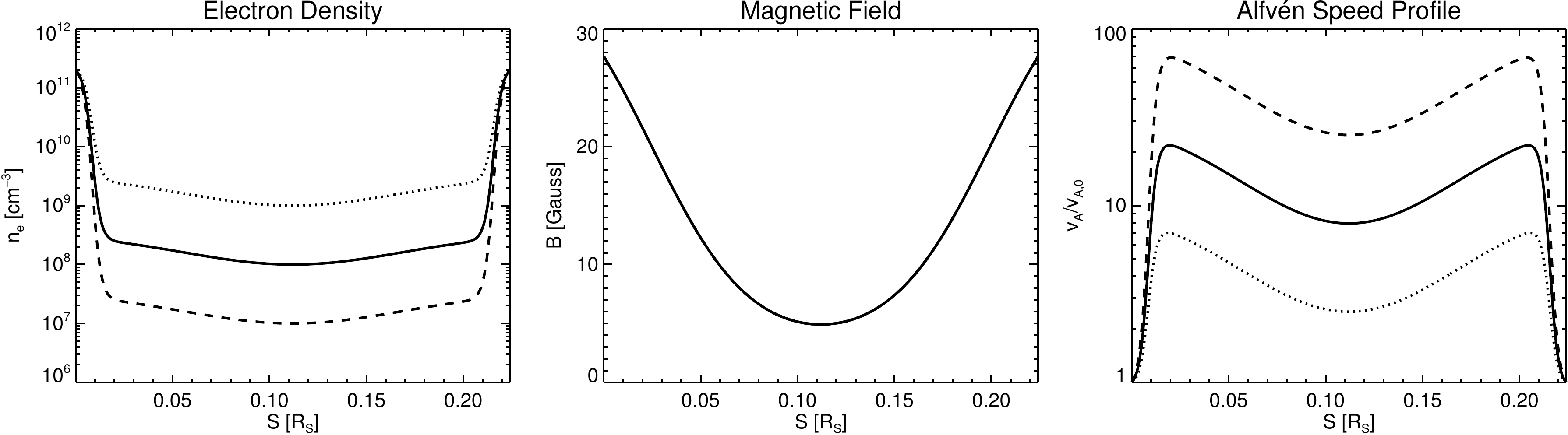}
\caption{Illustration of the symmetric analytic loop backgrounds used to examine the properties of the wave evolutionary equations (\cref{eq:zpzm}). We show the analytic profiles of electron density (left), magnetic field (middle), and Alfv\'en speed (right). The solid line is represents a nominal coronal loop, while the dashed/dotted lines have smaller/larger apex densities and exhibit a greater/lesser change in Alfv\'en speed from the base to the apex.}
\label{fig:analytic_backgrounds} 
\end{figure*}

An analytically specified loop background for three different apex densities is shown in \cref{fig:analytic_backgrounds}. The loop is defined as a symmetric semi-circle with no parallel flow ($u=0$) and the magnetic field decreases by a factor of 5.6 from the base to the apex, implying moderate areal expansion.  The three loop profiles are a nominal loop with moderate apex density (solid line), small apex density (dashed line), and large apex density (dotted line). These $n_e$ and $B$ profiles are are intended for illustration purposes, and are used in this section only. 

As will become clear, a natural parameter for characterizing wave reflection is the profile of the Alfv\'en speed along the loop relative to the left footpoint, which we display as ${v_A}/{v_{A,0}}$. Looking at this profile in the top right panel we see how loops, particularly rarefied ones, can naturally exhibit large overall changes in their Alfv\'en speed profile from the chromosphere to corona.

We now proceed by looking at what happens to wave amplitude solutions when terms are successively activated in \cref{eq:zpzm}. For illustration purposes we drive waves from the left footpoint only by setting $z_{+,0}=z_0=15$ {\kms} and $z_{-,L}=0$. 

\subsection{WKB evolution}
\label{section:wkb}

\begin{figure*}[hbtp]
\centering
\includegraphics[width=\cdfullpage]{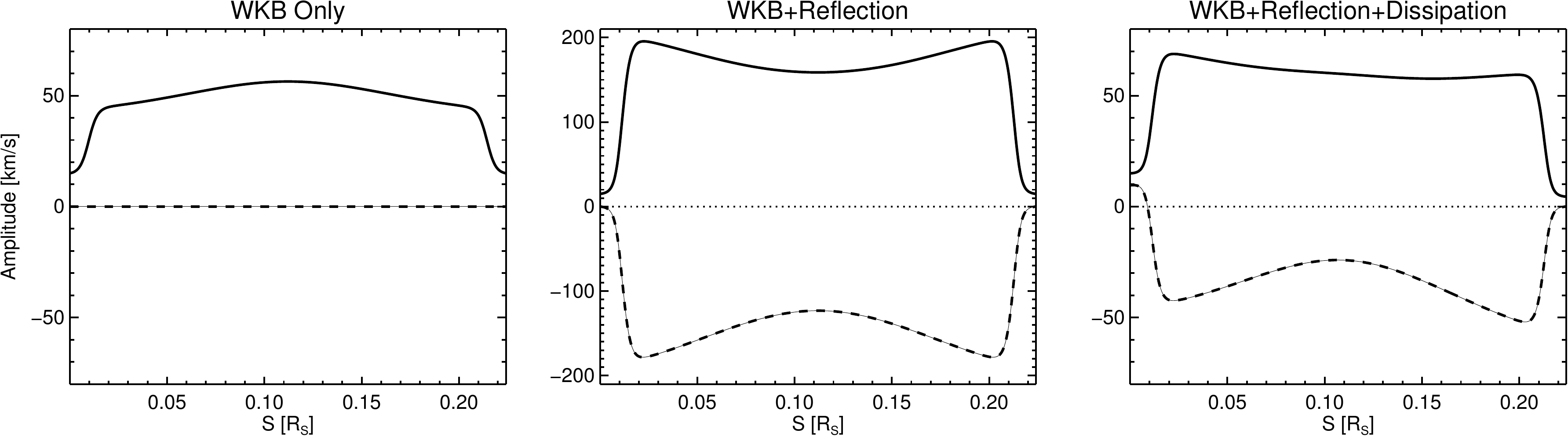}
\caption{Solutions of the wave evolutionary equations (\cref{eq:zpzm}) on the nominal loop background with successive terms activated. In each panel the $z_+$ amplitude is indicated by the thick solid line and the $z_-$ species is indicated by the thin solid/thick dashed line. Left: analytic solution for the wave amplitudes when only the WKB terms are activated.  Middle: analytic solution for the wave amplitudes when only the WKB and reflection terms are activated (note the change in scale). Right: numerical solution for the wave amplitudes when all terms activated. In each case waves are driven from the left footpoint only, making $z_+$ the dominant species. The minor species for these cases, $z_-$, is only generated when the reflection term is active.}
\label{fig:analytic_background_solutions} 
\end{figure*}

If we first disable the linear reflection and nonlinear dissipation terms (the last two terms on the RHS of \cref{eq:zpzm}), the equations become a simple recasting of the classic WKB equations in an amplitude form. $z_+$ has an analytic solution: $z_+(s) = z_0(\rho/\rho_0)^{-1/4}$, which is shown in the left panel of \cref{fig:analytic_background_solutions}. In this case $z_+$ propagates from left to right with no self-interaction and leaves the loop at the original amplitude on the other side. The profile of $z_+$ follows the density variation along the way, and the $z_-$ amplitude is identically zero because it was not driven from the right boundary. 

Because it will be useful in the subsequent discussion, we can also look at this solution in terms of energy conservation. Defining the Poynting flux of a species as 
\begin{equation}
P_\pm = \pm e_\pm v_A = \pm\frac{B\rho^{\frac{1}{2}}z_\pm^2}{4\sqrt{4\pi}},
\label{eq:pflux}
\end{equation}
we can quantify the energy per unit time passing through a point in the loop by multiplying by the local loop area and then substituting the WKB solution into this equation. Knowing $A=A_0B/B_0$, we obtain
\begin{equation}
AP_+ = \frac{A_0B_0\rho_0^{\frac{1}{2}}z_0^2 }{4\sqrt{4\pi}}= \text{Const,}
\label{eq:energy_from_pflux}
\end{equation}
 seeing immediately that it is constant along the loop\footnote{If we don't have $u<<v_A$, as is the case for the solar wind, then the conserved quantity becomes the wave action \citep[c.f.][]{jacques77}.}. While not surprising, this makes it clear that the WKB terms simply propagate the energy content of the waves along the loop without modification. It also indicates that specifying the Poynting flux of the incoming wave at one height is equivalent to specifying it at any other height, provided that only the WKB terms are operating. This property enables us to use our term-limiting technique to specify an effective injection height (Section~\ref{section:effective_rho}).
 
\subsection{WKB + Linear Reflection}
\label{section:wkb+refl}

When the linear reflection ($R_2$) term is activated, the solution becomes more interesting because $z_+$ can generate oppositely propagating $z_-$ via reflection in the presence of Alfv\'en speed gradients. We see now that the distribution of $z_+$ in the middle panel of \cref{fig:analytic_background_solutions} is much more strongly peaked near the base of the corona (note the 2x change in y-axis scaling) and has the imprint of the Alfv\'en speed profile, which changes rapidly in the transition region. Although initially counterintuitive, we also observe that the $z_-$ amplitude vanishes at the left and right boundaries despite being noticeably present in the coronal portion of the loop. 

The behavior of the reflected wave can be understood by examining the analytic, time-independent, solution to this system (a coupled set of linear ODEs). This can be expressed in matrix form in terms of the Alfv\'en speed profile $R_A=v_A/v_{A,0}$ and magnetic field profile, $R_B=B/B_0$,
\begin{equation}
\begin{pmatrix}z_+\\ z_-\end{pmatrix} = \frac{z_0}{2R_B^{1/2}} \begin{pmatrix}1+R_A&1-R_A\\ 1-R_A&1+R_A\end{pmatrix}\begin{pmatrix}1\\ -\frac{1-R_{A_L}}{1+R_{A_L}}\end{pmatrix},
\label{eq:reflection_analytic}
\end{equation}
where $R_{A_L}$ is $R_A$ evaluated at the right footpoint. At the left footpoint, $R_A=R_B=1$ by definition, which gives a simple expression for the reflected wave amplitude at the left boundary: 
\begin{equation}
z_{-,0} = -z_0\frac{1-R_{A_L}}{1+R_{A_L}}.
\label{eq:reflection_analytic_at_s0}
\end{equation}
This expression is identically zero for the case of a symmetric loop ($R_{A_L} = 1$) and grows when asymmetries in the beginning and final Alfv\'en speed are present. It implies that time-independent solutions of the wave evolutionary equations can depend strongly on the integral properties of the loop and not just the local ones, since the mere presence of large Alfv\'en speed gradients is not enough to reflect upward directed flux back through the same boundary. As we show in the next section, this is relevant for quantifying the overall heat deposition, because any flux lost due to self-reflection will \emph{not} be available for conversion to heat through the dissipation term. 

\subsection{WKB + Reflection + Dissipation}
\label{section:wkb+refl+dissp}

With a sense of how reflection works on its own, we can now proceed to the full evolutionary equation by adding the last term in \cref{eq:zpzm}, the phenomenological dissipation term. The coupled equations become nonlinear and do not possess a generic analytic solution. However numerical solutions can be computed, and one is shown for the nominal static loop background using $\lambda_0=0.01\ \text{R}_s$ in the right panel of \cref{fig:analytic_background_solutions}. Comparing to the case without dissipation, we see that the overall amplitude of $z_+$ is significantly reduced, and the profile is no longer symmetric from right to left due to dissipation along the way. Also noticeable is the fact that the reflected wave now has as a non-zero amplitude at the left boundary, meaning that input wave flux is now lost at the inner boundary through self-reflection, despite the fact that the overall Alfv\'en speed distribution is symmetric ($R_{A_L}=1)$.  

\begin{figure}[hbtp]
\centering
\includegraphics[width=\cdonecol]{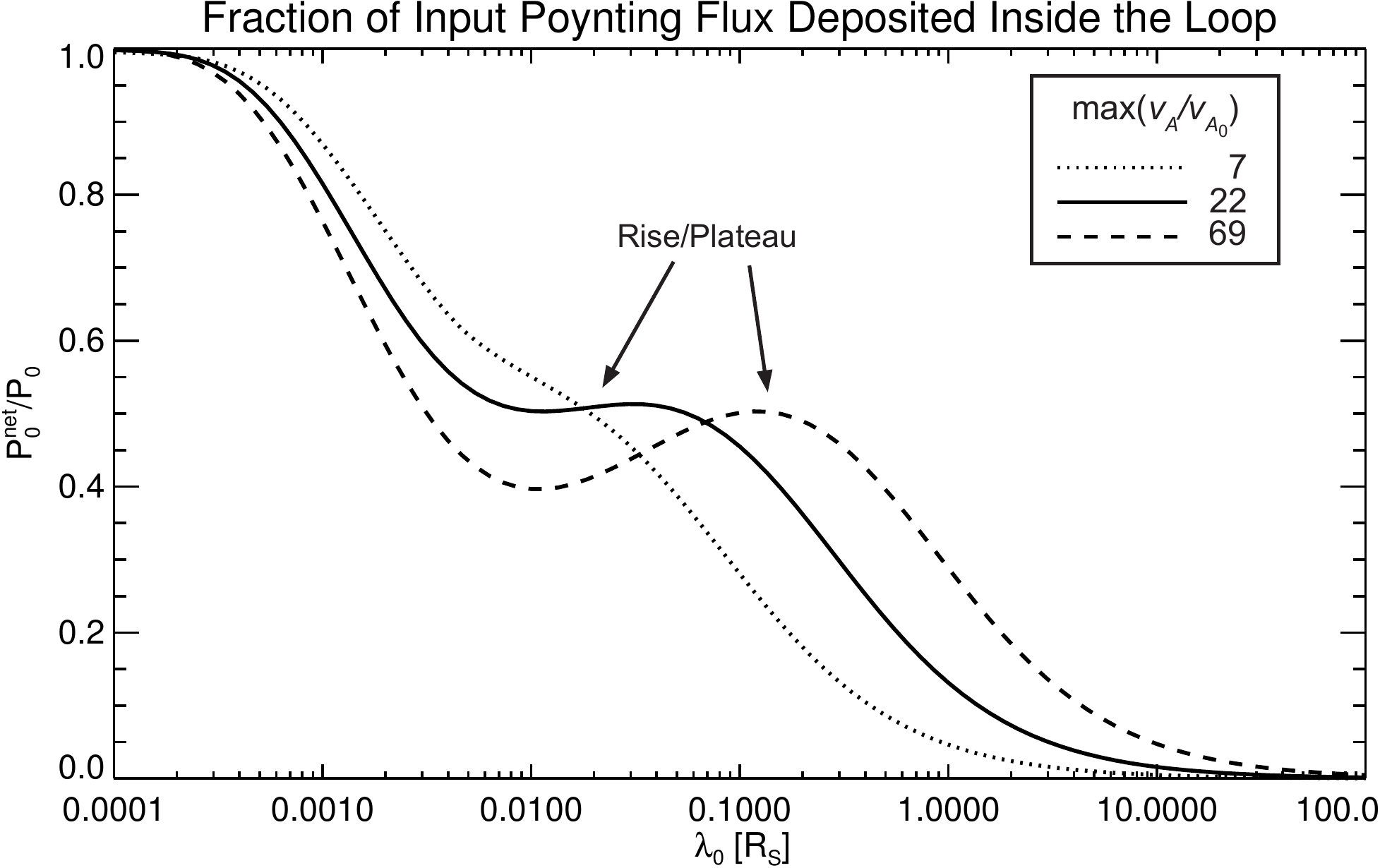}
\caption{The net heat flux deposited inside loop as a function of $\lambda_0$ for waves driven from one boundary only. The dotted, solid, and dashed lines show the results for loop backgrounds with increasing values of $\max({v_A}/{v_{A,0}})$. As the total change in Alfv\'en speed increases the role of self-reflection becomes more important, causing a rise or plateau in the center of the parameter space.
}
\label{fig:lambda_loss} 
\end{figure}

The newfound losses due to reflection suggest that dissipation, which continuously removes amplitude from $z_\pm$, prevents some of the flux from ``seeing'' the final Alfv\'en speed at the opposite end of the loop, giving an effective $R_A$ greater or less than one. Defining the net fraction of energy deposited in the loop in terms of the net Poynting flux at the boundaries,
\begin{equation}
\frac{P_{net}}{P_0} = \frac{P_{+,0} - P_{+,L}- P_{-,0}}{P_{+,0}} =  1-\frac{P_{+,L}+P_{-,0}}{P_{+,0}},
\end{equation}
we explore how this fraction varies as a function of $\lambda_0$ for all three Alfv\'en speed profiles in \cref{fig:lambda_loss}. $\lambda_0$  controls (inversely) the relative strength of the dissipation term to the other terms. For the smallest values of $\lambda_0$, dissipation dominates and all of the input flux is dissipated as soon as it is reflected locally. As dissipation becomes weaker, reflected wave energy can make it back to the left boundary and we see the net flux fraction begin to decrease to a local minimum followed by a short rise or plateau between $\lambda_0=0.01-1.0$. The location and depth of this plateau depends on the relative change in Alfv\'en speed from footpoint to apex, and becomes deeper for the case with a lower apex density (larger $v_A$). As dissipation becomes even weaker, flux is not dissipated enough before it reaches the opposite endpoint, at which point it is lost through the other side via transmission. 

\begin{figure}[hbtp]
\centering
\includegraphics[width=\cdonecol]{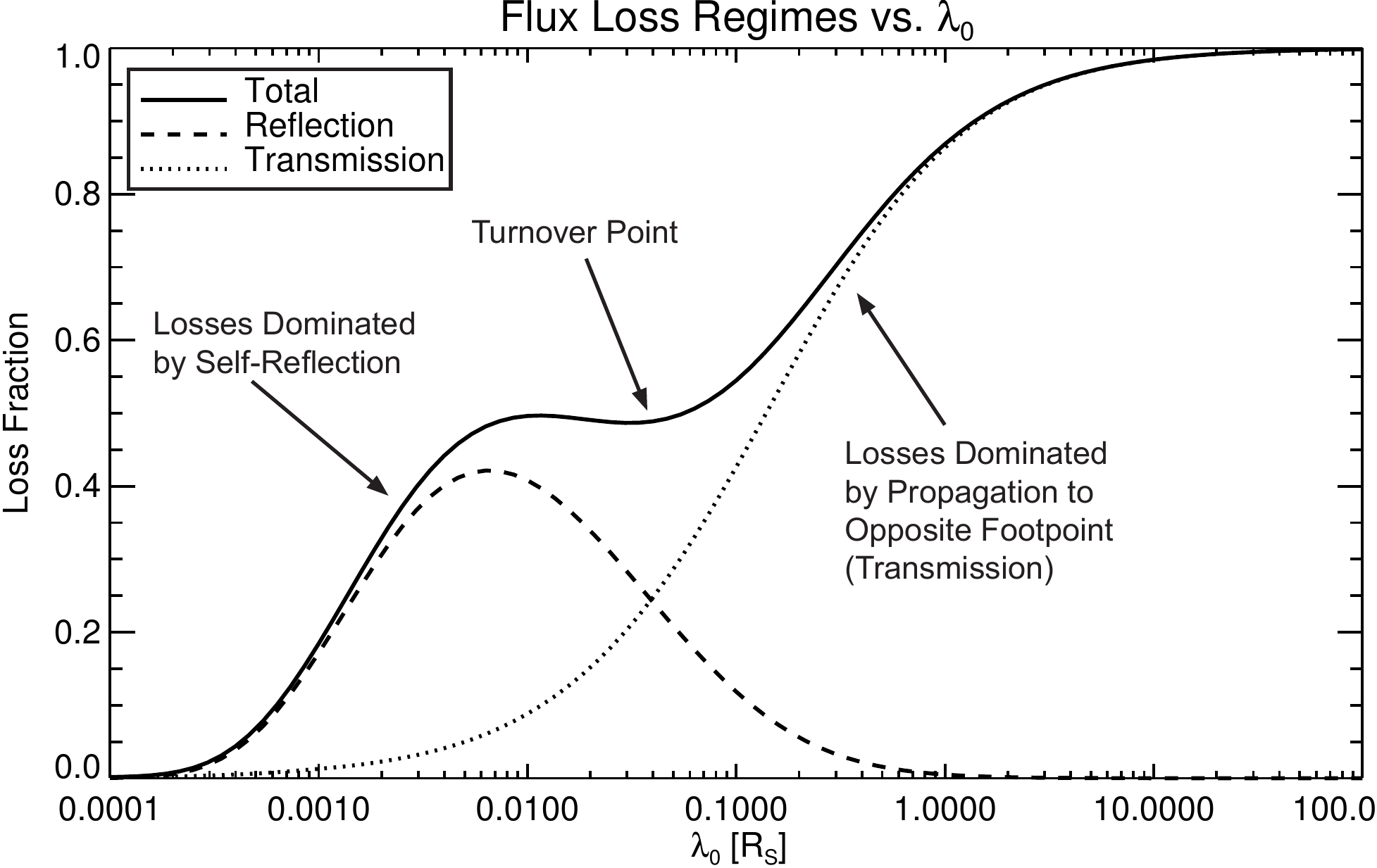}
\caption{Analysis of how the input Poynting flux is lost as a function of $\lambda_0$ for the nominal analytic loop background. The total loss curve (solid) is divided into the relative contributions of loss through the driven footpoint (reflection, dashed) and loss through the opposite footpoint (transmission, dotted). For small $\lambda_0$ (strong dissipation) the losses are dominated by self-reflection. For large lambda (weak dissipation) the losses are dominated by transmission.
}
\label{fig:lambda_contributions} 
\end{figure}

The relative contributions of loss through the left footpoint (reflection loss, $P_{-,0}/P_{+,0}$) and the opposite right footpoint (transmission loss, $P_{+,L}/P_{+,0}$) can be quantified by looking at the relative amount of flux leaving the domain on each side in \cref{fig:lambda_contributions}. Here we see that the reflection losses peak early, while the transmission losses dominate later, causing the conspicuous plateau that was seen in the net flux curves (\cref{fig:lambda_loss}).  In practice, the relative contribution of these loss terms will depend on the height and shape of the Alfv\'en speed profile along the loop as well as the relative strength of the dissipation term to the propagation terms.

\subsection{Driving Waves from Both Sides}
\label{section:analytic_both_sides}
In the previous section we examined the solutions for the case when the waves are driven from one side only. Not only was this a useful simplification for illustrative purposes, it enabled us to separate the behavior of the major (driven) from the minor (reflected) wave to produce \cref{fig:lambda_contributions}. In practice however, there will be wave flux driven from both sides of the loop, and the distinction between reflection and transmission becomes muddled in the two-wave formulation. 

\begin{figure}[hbtp]
\centering
\includegraphics[width=\cdonecol]{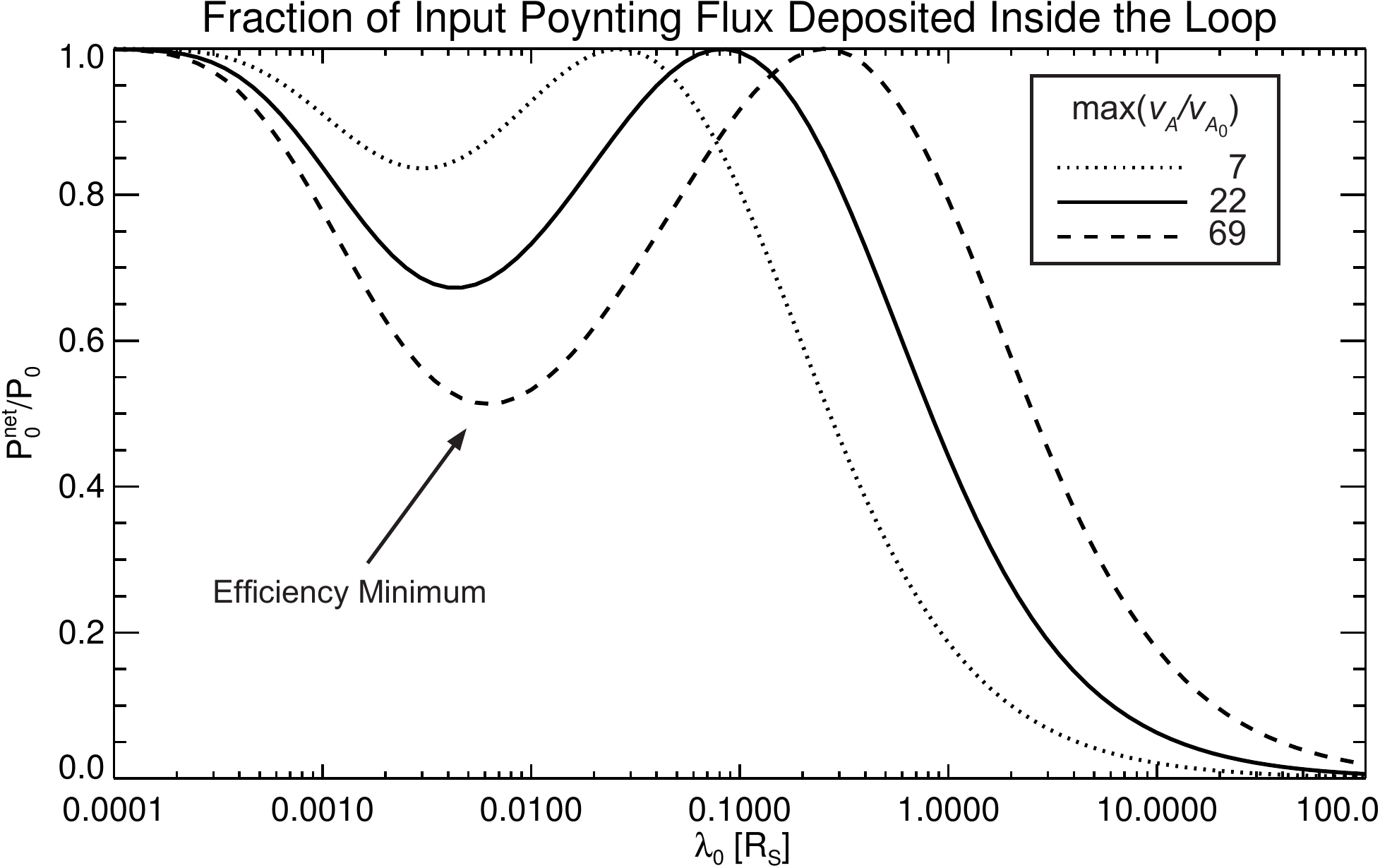}
\caption{Like \cref{fig:lambda_loss} but now for waves driven from both sides ($z_+$ on the left, $z_-$ on the right). Unlike the single-wave case, the fraction of deposited energy (efficiency) has a clear minimum and rise to a second peak at 1.0. The depth of the minimum and location of the second peak depend on the total change in ${v_A}/{v_{A,0}}$.
}
\label{fig:lambda_loss_both_sides} 
\end{figure}

To examine the key ramification of this change, we again calculate the net Poynting flux deposited as a function of $\lambda_0$ (like in \cref{fig:lambda_loss}), but using simulations that specify incoming wave flux from both sides ($z_{+,0}=z_0,\ z_{-,L}=-z_0$). The resulting loss curve is shown in \cref{fig:lambda_loss_both_sides}. Similar to the one-wave case, the curve again falls off when reflection becomes important, but instead of plateauing or peaking to some value less than one, the curve peaks at one again (i.e., full dissipation) before falling off in the transmission-dominated regime. This change results from the superposition of identical transmission/reflection behavior from both species feeding into one another, interchanging the roles of the dominant/minor species. The second peak occurs at the point when the amplitude of the minor species at each boundary crosses zero at the same time, setting the net losses to zero. In practice this perfect cancellation will only occur for symmetric loops, but this behavior is also relevant to loops with similar conditions at the left and right footpoints.

Looking at the change in net Poynting flux as the overall change in ${v_A}/{v_{A,0}}$ is increased, we observe that the first minimum shifts in location and gets increasingly deeper, reducing the efficiency (as before). We also observe that the second peak moves but always rises back to perfect efficiency before falling off again. This indicates that our choice of perfect correlation between the left driven and right driven species widens the overall efficiency regime of the model vs. a completely uncorrelated case (\cref{fig:lambda_loss}).

%--------------------------------------------------------------------------------------------------------------------------------------------
% Heating Analytic Calculations
%--------------------------------------------------------------------------------------------------------------------------------------------
\section{Analytic Properties of the Heating}
\label{section:heating_analytic}

\subsection{Energy Conservation}

To tie the preceding analysis back to coronal heating, we finish by examining the direct relationship between the net Poynting flux of the waves and the equivalent heating power per unit area, or heat flux, which can be defined as
\begin{equation}
H_F \equiv \frac{1}{A_T}\int_0^LAQ_{w}\ ds,
\label{eq:heatflux_definition}
\end{equation}
where $A_T=A_0+A_L$ is the total area of the footpoints.

A correspondence between the coronal heat flux and the wave fluxes is sensible from an energy conservation standpoint, but we can explicitly tie them together for the case when ($u<<v_A$) by multiplying \cref{eq:zpzm} by $A\rho z_\pm/2A_T$, summing the expressions for both species, and integrating over the loop with respect to $s$. The reflection terms cancel right away and the time-derivatives disappear for the steady-state assumption. The spatial derivatives and $R_1$ terms in the integrand for each species combine via the chain rule on the left hand side to give
\begin{equation}
\pm\frac{1}{4}\frac{A_0}{A_T}\frac{B_0}{\sqrt{4\pi}}\frac{\partial}{\partial s}\left (\rho^{1/2}z_\pm^2 \right ),
\label{eq:intermediate_poynting}
\end{equation}
which can be integrated between the loop endpoints. The remaining dissipation terms combine on the right hand side to look like the heating term times an area
\begin{equation}
-\frac{A}{A_T} \rho \frac{|z_-|z_+^2+|z_+|z_-^2}{4\lambda_\perp} = -\frac{A}{A_T} Q_W,
\label{eq:intermediate_dissipation}
\end{equation}
and the integral of this term is simply $-H_F$. Rearranging the integrated results, and assuming, by construction, that $\rho_0=\rho_L$, we arrive at an expression for the coronal heat flux in terms of $z_\pm$ at the left and right boundaries
\begin{equation}
H_F = \frac{1}{4}\frac{A_0}{A_T}\frac{\rho_0^{1/2}B_0}{\sqrt{4\pi}}\left [z^2_{+,0}-z^2_{+,L} - z^2_{-,0} + z^2_{-,L} \right ].
\label{eq:heatflux1}
\end{equation}
This expression can be further simplified and written in a source/sink form by rewriting $A_0/A_T$ as $B_L/(B_0+B_L)$ and assuming an equal driving amplitude for each driven (specified) species at the boundary $z_{+,0} =-z_{-,L}=z_0$, which gives us
\begin{equation}
H_F =\frac{1}{4}\frac{B_0B_L}{B_0+B_L}\frac{\rho_0^{1/2}}{\sqrt{4\pi}}\left [ \underbracket[1pt][0pt]{2z_0^2}_{sources}-\underbracket[1pt][0pt]{z^2_{+,L}-z^2_{-,0}}_{sinks} \right].
\label{eq:heatflux2}
\end{equation}
This form shows us that the heat flux into the loop is proportional to the harmonic mean of the footpoint fields and the driving amplitude squared. This is a powerful result because it immediately highlights a natural linear scaling of the wave heating model with the magnetic field, a necessary property for capturing a large dynamic range in the coronal heating rate\footnote{This relationship was pointed out in a more general discussion by \citet{sokolov13}.}. This form also highlights the importance of self-reflection and transmission in determining the final energy balance of the system (Sections \ref{section:wkb+refl} and \ref{section:wkb+refl+dissp}), because losses at the left and right boundary (sink) determine the fraction of input Poynting flux (source) that is actually deposited as heat.

Lastly, with some algebra it can be shown that $H_F$ is exactly equivalent to the average net Poynting flux passing through the loop boundary: 
\begin{equation}
P_{net} \equiv \frac{A_0(P_{+,0} - P_{-,0}) + A_L(P_{-,L} - P_{+,L})}{A_T}=H_F.
\label{eq:net_pflux_in_heatflux_terms}
\end{equation}
Using this expression we can define an efficiency ratio as the net Poynting flux to the input Poynting flux:
\begin{equation}
R_e \equiv  \frac{P_{net}}{P_{in}} = \frac{H_F}{P_{in}} = 1 - \frac{z_{+,L}^2 + z_{-,0}^2}{2z_0^2},
\label{eq:efficiency}
\end{equation}
which is a convenient way to show how wave energy losses at the ends of the loop reduce the total heat deposited.

\subsection{Heating Scale Height}
We complete the analytic discussion by exploring the dependence of the heating scale height on loop parameters. The relative steepness or stratification of the heating profile can play a key role in determining the hydrostatic solution (or lack thereof) for a given set of loop parameters \citep[e.g.][]{serio81,aschwanden02,mikic13}. For example, \citet{aschwanden01:trace} found that for a given apex temperature, hydrostatic solutions with heating concentrated near their footpoints required a larger total heat flux than their uniformly heated counterparts.  They also illustrated how when the scale height became too short for a given loop length, stable hydrostatic solutions cease to exist, suggesting that such loops would naturally exhibit dynamic evolution. 

Although the true $z_\pm$ solution depends on the nonlinear dissipation term and two-way coupling to the hydrodynamic solution, we can get a sense of the heating dependence by returning to the WKB+Reflection solution in \cref{eq:reflection_analytic}. For a loop that is driven from both sides and has identical footpoint conditions, the linear superposition of the left and right $z_\pm$ solutions gives a simple expression for the wave amplitudes:
\begin{equation}
z_\pm=\pm\ z_0\sqrt{\frac{R_A}{R_B}} = \pm\ z_0\sqrt{\frac{R_B}{R_\rho}},\\
\end{equation}
where $R_\rho = \rho/\rho_0$. This can be substituted into the heating term to determine the proportional dependence of $Q_W$ on loop parameters:
\begin{equation}
Q_W \propto R_BR_A = R_B^2R_\rho^{-1/2}.
\end{equation}
Now if the $Q_W$, $B$, $\rho$ profiles are locally approximated by exponential functions of height with the form $Ce^{-h/\lambda}$, then the exponents are related by:
\begin{equation}
\frac{1}{\lambda_Q} = \frac{2}{\lambda_B} - \frac{1}{2\lambda_\rho}, 
\end{equation}
where $\lambda_Q$, $\lambda_B$, $\lambda_\rho$ are the respective scale heights of each profile,  which gives a simple expression for the $\lambda_Q$ in terms of the other scale heights:
\begin{equation}
\lambda_Q = \frac{2\lambda_\rho\lambda_B}{4\lambda_\rho-\lambda_B}.
\label{eq:heating_scale_analytic}
\end{equation}
This is a key point because it indicates how the WTD heating profile will naturally adapt to specific conditions along a loop. The dependence on $\lambda_B$ immediately tells us to expect a steeper solution for rapidly expanding loops than nearly uniform ones, and the harmonic mean nature of the equation ensures there are regimes where either scale height can be relevant (though $\lambda_B$ has more weight). We derived this expression for the regime where the dissipation term is negligible with respect to the WKB and reflection terms, but it turns out to be surprisingly useful for a considerable range of the parameter space that we explore. This is demonstrated in Section~\ref{section:paramspace}.

%--------------------------------------------------------------------------------------------------------------------------------------------
% Single Loop Examples
%--------------------------------------------------------------------------------------------------------------------------------------------
\section{Solution Properties for a Single Loop}

\label{section:single_loop}
In this section we illustrate the behavior of the full system of coupled wave/hydrodynamic equations for a single loop. We use the same geometry as the perturbed semi-circular loop from section 5 of \citet{mikic13}, which has a loop length of $L=156$ Mm, and an apex that is slightly shifted to the right of the loop midpoint, occurring at $s=0.54L$. We similarly specify the magnetic field profile as $B(s)=B_0 f(s)^n$, where
\begin{equation}
f(s) = \frac{1}{11} + \frac{10}{11}\left (e^{-s/\lambda} + e^{-(L-s)/\lambda} \right ),
\end{equation}
and $\lambda=14$ Mm. Instead of $n=1$, we choose $n=\ln{5.5}/\ln{10.2}=0.73$, which gives a smaller total areal expansion (5.5 instead of 10.2) and a shallower expansion overall. For this loop we study two cases, Case A has a moderate footpoint magnetic field of $B_0=55$ Gauss, and Case B has a stronger field of $B_0=200$ Gauss. For each case, we set a driving amplitude of $z_0=5.2$ {\kms} at the loop footpoints. This corresponds to an effective amplitude of $12$ {\kms} at $n_e^{\text{eff}}=$\ \neu{2}{11}{\un} (Sections~\ref{section:effective_rho} \& \ref{section:wkb}). The full system of equations is integrated for over 100 sound crossing times, corresponding to 700+ Alfv\'en crossing times at a minimum and about 40 hours of physical time.

\subsection{Case A}
\label{section:single_loop_caseA}

\begin{figure*}[hbtp]
\centering
\includegraphics[width=\cdfullpage]{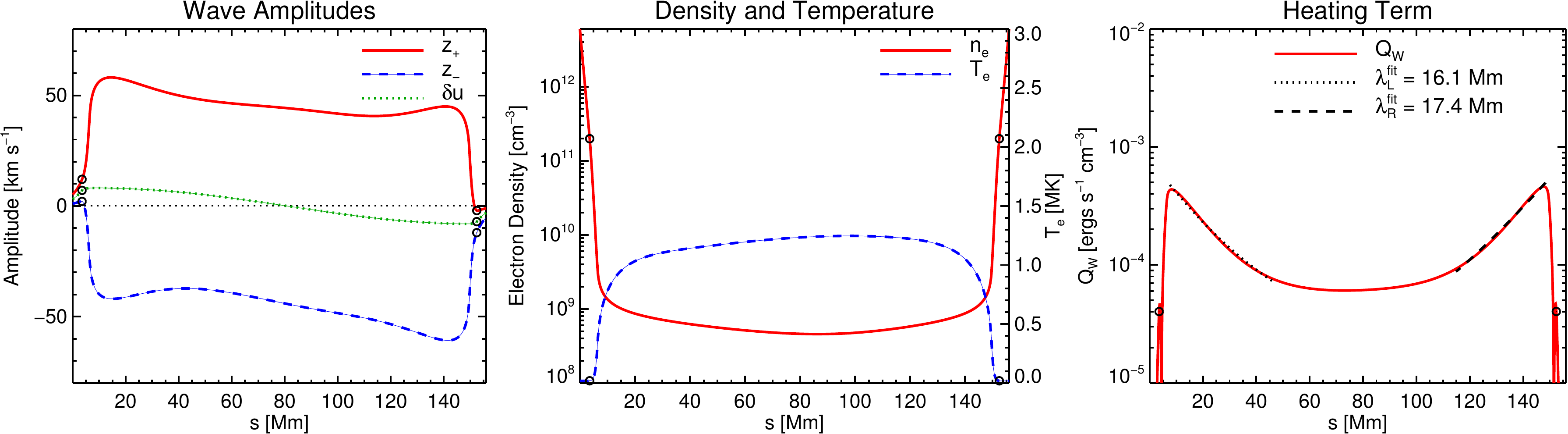}
\caption{Summary of the steady-state solution for Case A as a function of the loop coordinate. The left panel shows the wave amplitudes for the $z_+$ (red) and $z_-$ (blue) species, which are driven from the left and right footpoint respectively. $\delta_u$, determined from $z_+$ and $z_-$, is shown in green. The middle panel shows the electron density (red) and temperature (blue), which are slightly asymmetric due to the perturbed loop geometry. The right panel shows the heating rate from the wave dissipation term. Fits to the profile on the left and right sides of the loop are indicated in the dotted and dashed black lines respectively. The location of the transition to the complete $z_\pm$ equation at $n_e^{\text{eff}}$ is indicated by the circles in each trace (Section~\ref{section:effective_rho}).
}
\label{fig:single_summary_caseA} 
\end{figure*}

\Cref{fig:single_summary_caseA} shows the steady-state solution that was achieved for Case A. The $z_\pm$ traces in the left panel show clear dissipation and reflection signatures, peaking just above the transition region and decreasing from left to right and right to left for the + and - species respectively. The point at which we transition to the full non-WKB terms below $n_e^{\text{eff}}$ is indicated by the open circles overlaid on each trace. It is evident at these points that the amplitude of the driven species dominates the amplitude of the oppositely propagating wave (i.e., the net Poynting flux). Assuming a perfect correlation of the wave species, we can also estimate the average velocity fluctuation from the Els\"asser variables at each point along the loop: $\delta u=(z_+ + z_-)/2$. For this case, the maximum value of $\delta u$ is 8.2 {\kms}, and $\delta u=7.7$ {\kms} at a low coronal height of $h=21$ Mm. Averaging the absolute value over the whole loop gives a mean value of $|\delta u|$ of 5.3 {\kms}.

The middle panel of \cref{fig:single_summary_caseA} shows the steady-state density and temperature profiles. These profiles are slightly asymmetric due to the perturbation of the apex position and show a modest apex density of \neu{4.6}{8}{\un} and maximum temperature of 1.25 MK. The heating profile is shown in the right panel on a logarithmic scale. This profile is clearly stratified, having a maximum of  \mbox{$\sim\, $\neu{5}{-4}{\uh}} at the coronal base followed by a rapid decay to a floor of \mbox{$\sim\,$\neu{6}{-5}{\uh}} at the apex. In integral terms, the input Poynting flux of the wave energies is \neu{3.3}{6}{\uf} while the heat flux is \neu{3.16}{6}{\uf}. This gives $R_e\!=\!0.96$, which indicates that most of the wave energy is deposited inside the loop for these parameters.

We can characterize the relative stratification of the heating by fitting the profile on each side to an exponential decay function, $Q=Q_0\exp(-h/\lambda_Q)$, where $h$ is the height above the solar surface, and $\lambda_Q$ is the corresponding heating scale height. For these fits (and all subsequent scale height fits described in the paper) we fit the profile for all points above the left or right footpoint that lie between 15\% and 75\% of the maximum loop height and are at coronal temperatures or above ($T_e\ge0.4$MK). The fits to the left and right footpoints are shown as the dotted and dashed lines over the heating profile, and the corresponding scale heights are indicated. These relatively short scale heights are between a 4.5-5 times smaller than the loop half-length (76 Mm) and indicate that the heating from the WTD driven model can be relatively concentrated at the footpoint.  

\subsection{Case B}
\label{section:single:caseB}

The loop solution becomes even more interesting when we increase the basal magnetic field from 55 to 200 G while keeping all other parameters the same. This increases the input Poynting flux by a factor of 3.6 and will cause a similar increase in $H_F$ provided there are no drastic changes to $R_e$.  With this strong increase in overall heating, the solution can no longer reach a time-independent steady-state solution and instead exhibits a repeating heating and cooling cycle with a period of 4.6 hours. 

\begin{figure*}[hbtp]
\centering
\includegraphics[width=\cdfullpage]{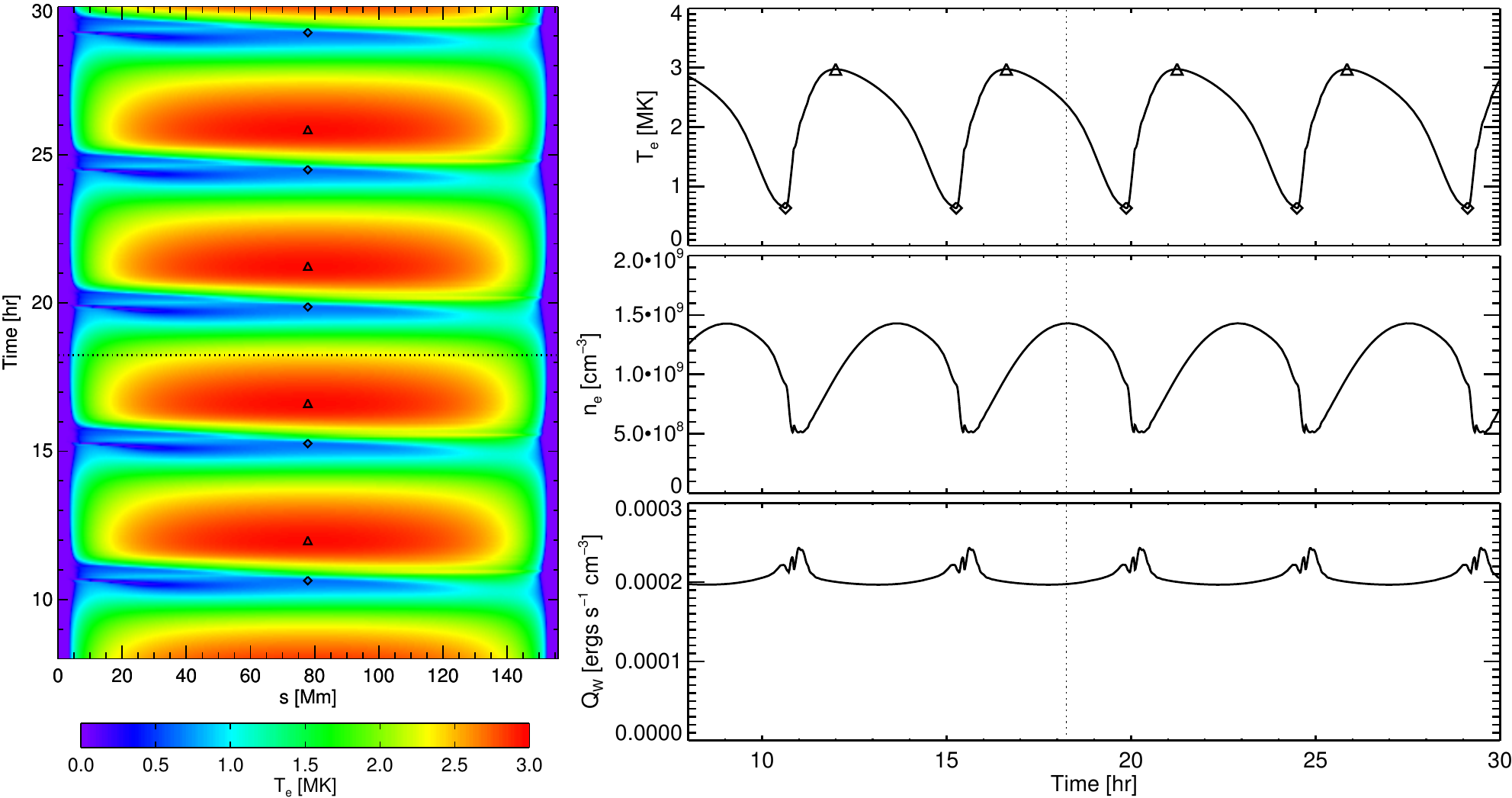}
\caption{Illustration of the thermal nonequilibrium cycles from Case B. The left panel shows a 2D color plot of $T_e$ as a function of the loop coordinate and time. Identical cycles of heating and cooling occur, with an incomplete condensation forming and eventually exiting the loop on the left side each time. The right panels show the temperature, density, and heating rate from top to bottom as a function of time at the loop midpoint. The triangles and diamonds indicate the maximum and minimum temperatures of the cycle respectively. 
}
\label{fig:single_tne} 
\end{figure*}

The time evolution of the cycle is illustrated in \cref{fig:single_tne}, which shows a two dimensional plot of temperature as a function of loop position and time. The right hand side of the figure shows traces of $T_e$, $n_e$, and $Q_w$ as a function of time at the midpoint of the loop. This is quite similar to the thermal nonequilibrium (TNE) cycles describe by \citet{mikic13}. TNE cycles involve a long cooling phase driven by sustained force imbalance, which increases density and lowers temperature near the loop top. This configuration, aided by the lack of perfect symmetry, becomes unstable before a complete condensation forms, causing the cooling material to collapse to one side and be removed from the loop. The rarefied loop then heats up rapidly and begins the cooling phase again shortly thereafter. We can quantify the TNE cycle for this loop by finding the repeated minima and maxima of temperature (triangles and diamonds in \cref{fig:single_tne}) and computing the period, $\tau=4.6$ hrs, maximum temperature, $T_{max}=2.97$ MK, minimum temperature, $T_{min}=0.64$ MK, and the midpoint temperature measured at the temporal midpoint of the cooling phase, $T_{mid}=2.38$ MK (dotted line).

Looking at the time evolution of the heating rate we see that while the temperature and density undergo large oscillations during the TNE cycle, the relative variation of the heating rate is much smaller, especially during the cooling phase when the density is grows tremendously. This is a striking result because it illustrates how the WTD heating rate can be relatively \emph{insensitive} to evolution of the hydrodynamic state of the loop, even when temperatures and densities are changing considerably during the TNE cycle. 

\begin{figure*}[hbtp]
\centering
\includegraphics[width=\cdfullpage]{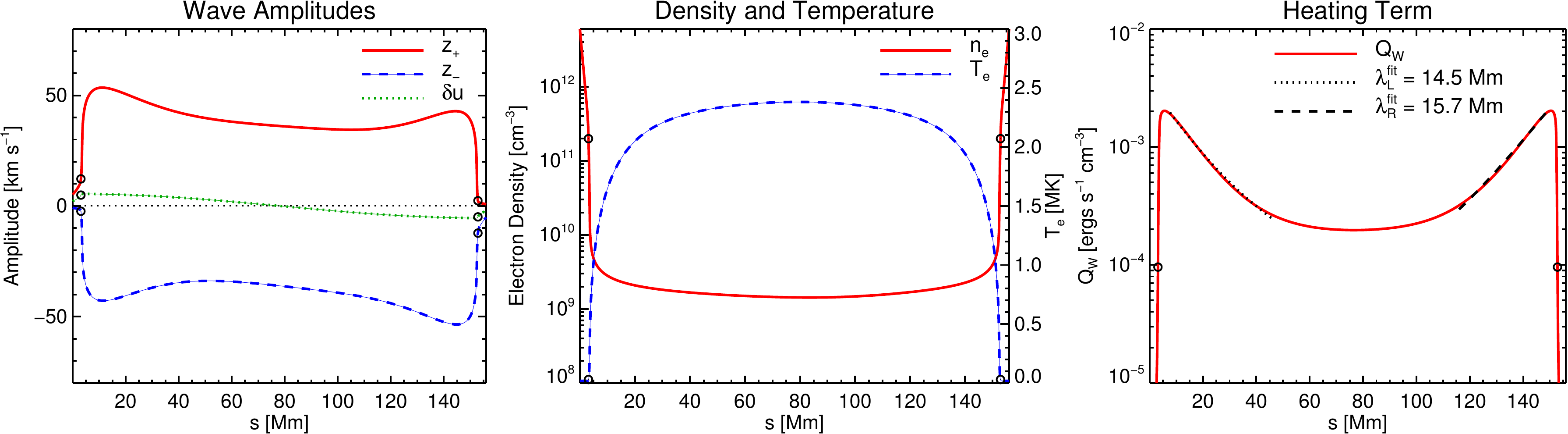}
\caption{Same as \cref{fig:single_summary_caseA} but now for Case B. This solution exhibited non-steady TNE behavior, and we show the solution at the temporal midpoint of the cooling cycle. The increased field strength imposes an input Poynting flux that is almost a factor of four larger than Case A. This results in larger heating rates, densities, and temperatures while the shape of the heating profile is quite similar.
}
\label{fig:single_summary_caseB} 
\end{figure*}

Returning now to contrast the two cases, we show traces of $z_\pm$, $n_e$, $T_e$, and $Q_w$ at the midpoint in time of the TNE cooling cycle for Case B in \cref{fig:single_summary_caseB}. The $z_\pm$ profiles are quite similar in size and shape despite the noticeable temperature and density disparity between the two cases. The resulting heating profile is also similar in shape, but is larger by about a factor of four, as expected.

%--------------------------------------------------------------------------------------------------------------------------------------------
% Paramspace Study
%--------------------------------------------------------------------------------------------------------------------------------------------
\section{Parameter-Space Exploration}
\label{section:paramspace}

With example solutions to the coupled hydrodynamic/wave system covered in the previous section, we now proceed with a parameter-space exploration of the model. The major goal of these experiments is to characterize the performance and scaling of the model for a broad range of possible conditions. Understanding the basic scaling properties of the model is key to establishing for what regimes (if any) it can be a candidate coronal heating mechanism. 

This exercise also helps us determine what loop parameters influence the shape and amplitude of the heating profile. This is important because this heating formulation, unlike those from purely empirical models, only has two real free parameters ($z_0$, $\lambda_0$), and the rest are set implicitly by the magnetic field strength, loop geometry, and hydrodynamic evolution. Although this lack of freedom is a desirable trait, it does require the formulation to naturally adapt to the myriad of solar conditions that it will encounter.

Using Case A from the previous section as our reference run, we conduct a simple parameter-space study by varying important parameters one at a time, and running 20-22 simulations for each case (104 total). The five parameters that we chose can be divided into two distinct groups. The first group consists of parameters set by the background structure of the loop: the base magnetic field strength, $B_0$, the maximum loop expansion ratio, $\Gamma=\max(A/A_0)$, and the total loop length, $L$. The second group contains the model's two free parameters: the input wave energy, $e_0 = \rho_0z_0^2/4$, and the base correlation length, $\lambda_0$. The important distinction between these groups is that the former are set by the background structure of the corona (i.e., determined by the problem at hand), while the latter can be varied independently.

Because of the abundance of simulation results and various ways in which they can be displayed, we narrow our discussion to the most salient heating features and scaling properties. For these purposes we examine the heat flux, $H_F$ (Eq. \ref{eq:heatflux_definition}), the apex temperature, $T_A$, and the effective scale height of the heating, $\lambda_Q$.  Since the heating profile may be asymmetric, $\lambda_Q$ is defined as the arithmetic mean of the left and right exponential fits to heating profile.

\subsection{Magnetic Field Strength}

\begin{figure*}[hbtp]
\centering
\includegraphics[width=\cdfullpage]{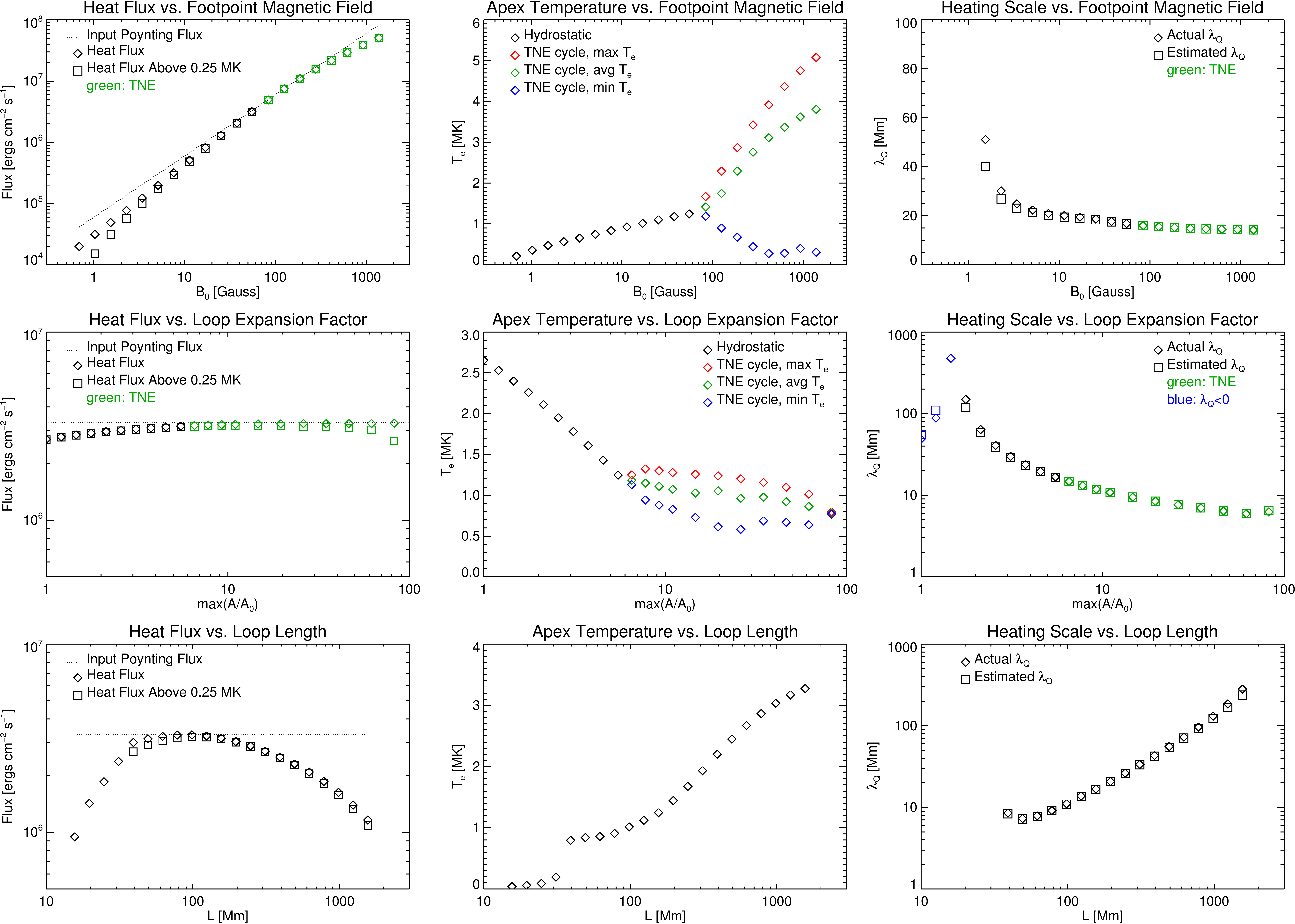}
\caption{Parameter-space results for the variables that depend on the pre-existing loop geometry. The x-axis of the rows show the footpoint magnetic field strength (top row), maximum areal expansion (middle row), and total loop length (bottom row). From left to right we show the total heat flux deposited (left column), the apex temperature (middle column), and the fitted heating scale height (right column) in diamonds as a function of these parameters. Green coloring is used to indicate loops that underwent TNE cycles. For these cases, the min $T_e$, max $T_e$, and midpoint $T_e$ of the cooling phase are indicated in the middle column. Boxes in the left column indicate the total heat flux deposited above 0.25 MK and the dotted line indicates the input Poynting flux (i.e., no losses). Boxes in the right column indicate the estimated heating scale height from the density and magnetic field scale heights (Eq. \ref{eq:heating_scale_analytic}).
}
\label{fig:pspace_loop_params} 
\end{figure*}

The results for the three loop parameters are shown in \cref{fig:pspace_loop_params}, and we begin by examining the dependence on the base magnetic field strength in the top row. $B_0$ is varied from 0.7 to 1400 G, and this span is covered by 20 simulations in log space. The left column shows the variation of $H_F$ with $B_0$ (diamonds) along with the input Poynting flux (dotted line), which is linearly proportional to $B_0$. We immediately we see that $H_F$ has a near linear scaling of with $B_0$, especially between 10-500 G, and that $H_F$ begins to depart from perfect scaling at both the low and high ends of of the range.  This can be understood as a change in the relative efficiency ($R_e$, \cref{eq:efficiency}) of the model as it transitions from a net Poynting flux with losses dominated by self-reflection for small $B_0$, to one with losses dominated by transmission for large $B_0$.

The effect of this scaling is made tangible by looking at $T_A$ vs $B_0$ in the middle column. The apex temperature starts very low due to the small energy input for weak $B_0$, and slowly rises as $H_F$ grows with $B_0$. At a certain point steady-state solutions are no longer possible and the hydrodynamic solutions exhibit TNE cycles. When this occurs we plot 3 symbols for each run, indicating the maximum, minimum, and midpoint temperatures of the cycle in red, blue, and green respectively. These numbers are calculated in the same manner as illustrated for Case B in Section~\ref{section:single:caseB}. Most notably, this transition marks a steep increase in the scaling of temperature and indicates that high active-region like temperatures and time-dependent evolution occur naturally for reasonable values of $B_0$.

In the rightmost column, we show how the effective heating scale height varies with $B_0$. Our fitted $\lambda_Q$ is undefined due to low temperatures for the two leftmost runs, but we otherwise see a rapid drop from a moderate 50+ Mm height to a nearly constant 15-20Mm range between $B_0$ 10-1000 G. That $\lambda_Q$ decreases by only 25\% or so over two orders of magnitude past $B_0=10$ G  indicates that the heating scale height is not strongly dependent on the field magnitude once coronal solutions ($T_A > $1.0 MK) are established. These lengths are also relatively short for the loop ($L/2=78$ Mm), which is consistent with the appearance of non-hydrostatic TNE solutions once the heating power becomes large enough.

\subsection{Areal Expansion}

Next we want to characterize how loop expansion influences the heating profile. For realistic magnetic field configurations there are myriad ways in which the magnetic field (and thus the cross-sectional area) can vary along the loop. However, we can get a basic handle on this parameter by scaling the reference loop area profile to achieve a range of expansion factors. This is done by defining the new magnetic field profile of the loop as:
\begin{equation}
B_\Gamma = B_0\left ( \frac{B}{B_0} \right )^{\ln \Gamma/\ln \Gamma_0},
\label{eg:area_paramspace}
\end{equation}
where $B$ and $B_0$ are the magnetic field profile and basal field strength, and $\Gamma_0$ and $\Gamma$ are the expansion factor of the reference loop and the desired areal expansion factor.

Using this method we explore a range of areal expansion factors from $\Gamma\!=\!1$ (uniform loop) to $\Gamma\!=\!110$ (rapidly expanding) and these results are shown in the middle row of \cref{fig:pspace_loop_params}. First we see that the heat flux deposited does not vary much, with only a slight dip in $R_e$ for small $\Gamma$. This relative flatness is expected because the input Poynting flux is constant for these runs. What we do notice however is a stark drop in the apex temperatures from a relatively high 2.65 MK for $\Gamma\!=\!1$ to 1.25 MK for the reference run at $\Gamma\!=\!5.5$. Beyond this point, a threshold is reached and non-hydrostatic TNE solutions occur, and the midpoint temperature of the TNE solutions follows a shallower slope than before. 

Two things are in play in determining the temperature behavior. The most obvious factor is the scale height (right column), which shows a clear decreasing trend on a log scale out to $\Gamma\!=\!82$. As discussed by \citet{aschwanden01:trace}, more heating power is required to maintain a fixed $T_A$ when $\lambda_Q$ decreases. Therefore when $H_F$ is fixed, a decreasing $T_A$ is expected. At a certain point $\lambda_Q$ will become too small with respect to $L$ to support hydrostatic solutions, and indeed we see that TNE sets in around $\lambda_Q\!=15$ Mm (about $5\times$ less than $L/2$). The physical reason that $\lambda_Q$ drops with $\Gamma$ is due to the dependence of $z_\pm$ on the variation of the magnetic profile along the loop, and the drop nearly exactly follows the analytic approximation to $\lambda_Q$ given in \cref{eq:heating_scale_analytic}. Here $\lambda_\rho$ changes very little, but $\lambda_B$ naturally decreases as $\Gamma$ increases and the $B$ profile steepens.

The second factor is due to the geometry of an expanding loop. Because $H_F$ is fixed for these runs, when the area factor in \cref{eq:heatflux_definition} increases then $Q_w$ must decrease in such a way that the integral remains constant. In other words, if the area at the base of the loop is fixed, then the expanding loop must deposit the same total energy per unit time over a larger relative volume than a uniform loop. This will lower the average heating rate and therefore contribute to the drop in $T_A$ with $\Gamma$.

Overall we see that the heating scale height and heating rate of the WTD model are intimately tied to the expansion factor. This is a desirable trait because it shows that the heating naturally adapts to the individual magnetic profile of a loop.

\subsection{Loop Length}

The other primary geometric parameter is the loop length, $L$. $L$ is a key parameter in coronal heating scaling laws because it directly influences the amount of heating power required to maintain a given apex temperature---shorter loops require more heat than longer ones \citep{rosner78,aschwanden02}. To explore the dependence on $L$, we fix all other parameters describing the reference loop and multiply length scales by a constant factor, covering a logarithmic range of $L$ from 16--1600 Mm. This scaling is somewhat artificial because the areal expansion profile and magnetic field strengths will change with $L$ for realistic fields, but it allows us to separate out the $L$ dependence.

The bottom row of \cref{fig:pspace_loop_params} shows the dependence of $H_F$, $T_A$, and $\lambda_Q$ on $L$. The first few solutions are simply too short to achieve coronal temperatures given the relatively small $H_F$, but afterwards we see a clear increasing trend of $T_A$ with $L$ (middle column).  This increasing trend is consistent with coronal heating scaling laws, but the details are influenced by the growing flux losses at both ends of the $L$ range (left column). On the short end, the loops are too short for appreciable dissipation, and wave flux is lost by transmission when it reaches the opposite side of the loop. On the long end, the Alfv\'en speed increases as the average density drops, causing the relative $v_A$ profile to grow, causing self-reflection losses to become a factor.

We also see that the scale height correlates well with $L$ (right column). This is not surprising because the magnetic field profile was scaled with $L$ by construction, however it reinforces the notion that the WTD model will naturally adapt the heating profile to a given loop geometry.

\subsection{Wave Energy}
We next turn to the tunable, or free parameters of the WTD heating model. The most obvious parameter is driving wave energy at the boundary, $e_0$. Like $B_0$, $e_0$ controls the input Poynting flux of waves into the domain, and we examine loop solutions that span over four orders of magnitude in input flux. Since $\rho_0$ is fixed in our runs, $e_0$ is entirely controlled by varying the driving amplitude, $z_0$ (see \cref{eq:ew}).

\begin{figure*}[hbtp]
\centering
\includegraphics[width=\cdfullpage]{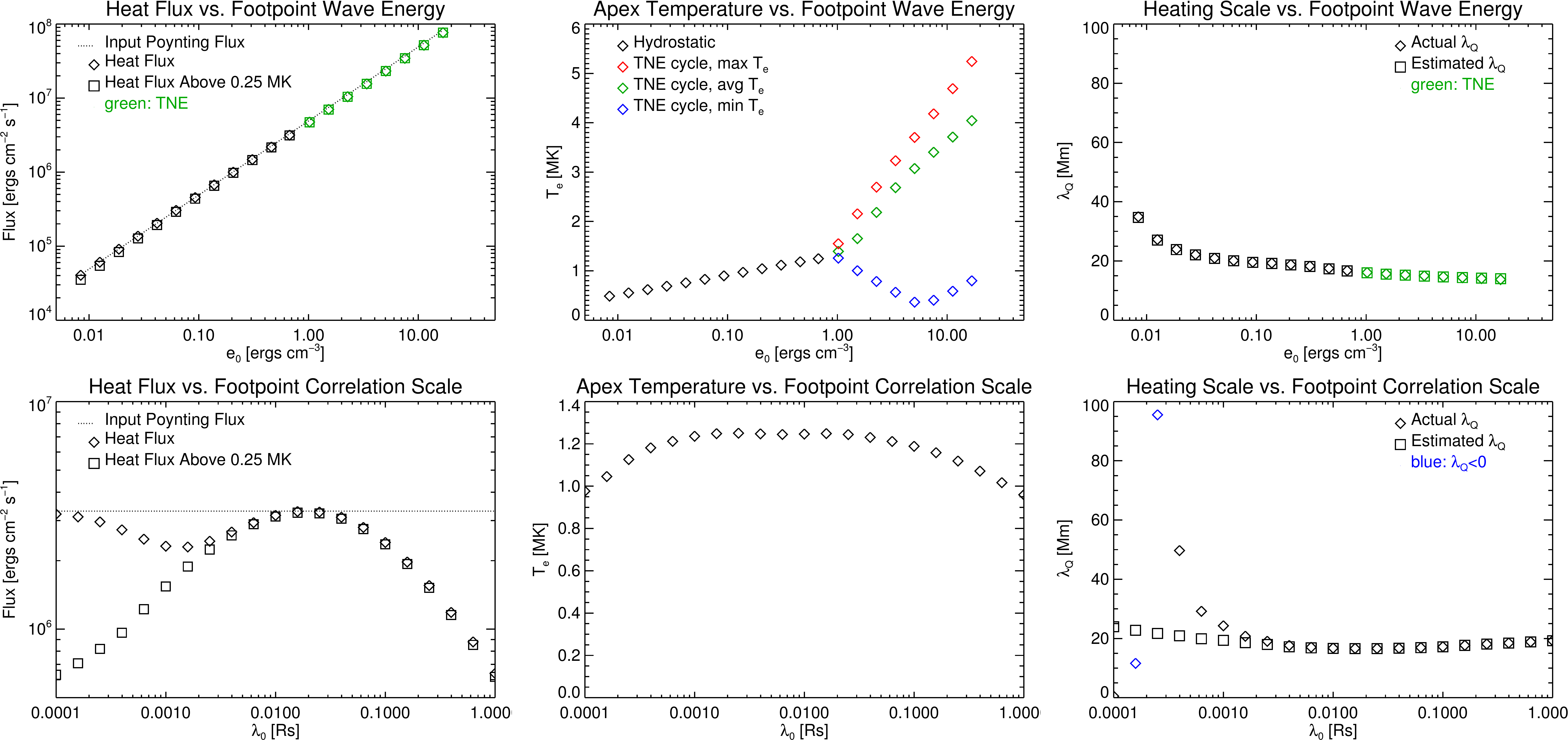}
\caption{Same as \cref{fig:pspace_loop_params} but now for the two independent parameters of the model, the footpoint wave energy (top row) and the correlation length of turbulence at the footpoint (bottom row).
}
\label{fig:pspace_free_params} 
\end{figure*}

Looking at the top row of \cref{fig:pspace_free_params} we see a similar scaling of $H_F$, $T_A$ and $\lambda_Q$ compared to the $B_0$ runs, but with an increased efficiency fraction for small and large values of $e_0$. This is because the Alfv\'en speed and $\lambda_\perp$ do not also scale with $e_0$, causing a reversed but much shallower dependence of the efficiency fraction. 

It is also important to note that in reality $e_0$ is not an entirely free parameter because of observational constraints on the non-thermal motion present in coronal emission lines. The velocity fluctuations, $\delta u$, will scale with the square root of $e_0$, meaning that observations of non-thermal motions should present an upper limit to $e_0$ (and thus the Poynting flux) for a given magnetic field distribution. We explore how these fluctuations vary for all of the parameter-space runs in Section \ref{section:paramspace:deltau}.

\subsection{Perpendicular Scale Length}
The second tunable parameter is the perpendicular scale length constant, $\lambda_0$. As mentioned in of Section~\ref{section:wkb+refl+dissp}, $\lambda_0$ conveniently controls the relative strength of dissipation with respect to the other terms in the $z_\pm$ equations. In this case, we vary $\lambda_0$ from 0.0001 to 1.0 $R_S$ and examine the evolution of the full system as we transition from a dissipation-dominated (small $\lambda_0$) to a propagation-dominated regime (large $\lambda_0$).

The scaling results are shown in the bottom row of  \cref{fig:pspace_free_params}. The profile of $H_F$ (diamonds, left column) nicely mirrors the double-peaked profile of \cref{fig:lambda_loss_both_sides}. As discussed in Section~\ref{section:analytic_both_sides}, this indicates first a transition from total dissipation (no losses) to losses dominated by reflection ($\lambda_0\sim 0.001 R_S$). This regime is followed by another efficiency peak ($\lambda_0\sim 0.02 R_S$), after which losses become dominated by transmission. 

Unlike the analytic cases, the background solution is now coupled to $H_F$ and the heating distribution. For the smallest values of $\lambda_0$ only a small portion of the heat flux is actually deposited in the corona above $T_e=0.25$ MK (boxes). Here the dissipation term is so strong that the bulk of the energy is lost before it reaches the corona, and this leads to the drop in $T_A$ for small $\lambda_0$. 

Beyond the dissipation-dominated regime ($\lambda_0\gtrsim0.001\ R_S$) we see remarkably flat curves for $T_A$ and $\lambda_Q$ over nearly two orders of magnitude. This relative constancy is due to the relatively broad width of the net efficiency curve of as it transitions between loss regimes. Most importantly this result indicates that as long as $\lambda_0$ is chosen reasonably, the loop solutions are not strongly dependent on it. This is a desirable property for unifying this model with the heating and acceleration of the solar wind on open flux tubes, where the choice of $\lambda_0$ can play a more central role \citep{lionello14}.

\subsection{Implications for $\delta u$}
\label{section:paramspace:deltau}
With the heating properties of the five-variable parameter space explored, it is worth it to briefly examine the velocity fluctuations determined from the model, $\delta u = (z_+ + z_-)/2$. $\delta u$ is a relevant parameter because such motions can contribute to the non-thermal widths of spectral line profiles. Non-thermal widths constrain the amount of unresolved fluctuations present in the emitting plasma element, making them a useful observational constraint of wave heating in the corona \citep[e.g.][and references therein]{bemporad12,hahn14,brooks16}. Owing to the simplicity of our model, it does not make sense to quantitatively compare $\delta u$ to observations \citep[as more sophisticated models of Alfv\'enic turbulence have done, e.g., ][]{asgari-targhi14}, but this parameter can give us a qualitative sense of the range of velocity fluctuations present in the model.

\begin{figure*}[hbtp]
\centering
\includegraphics[width=\cdfullpage]{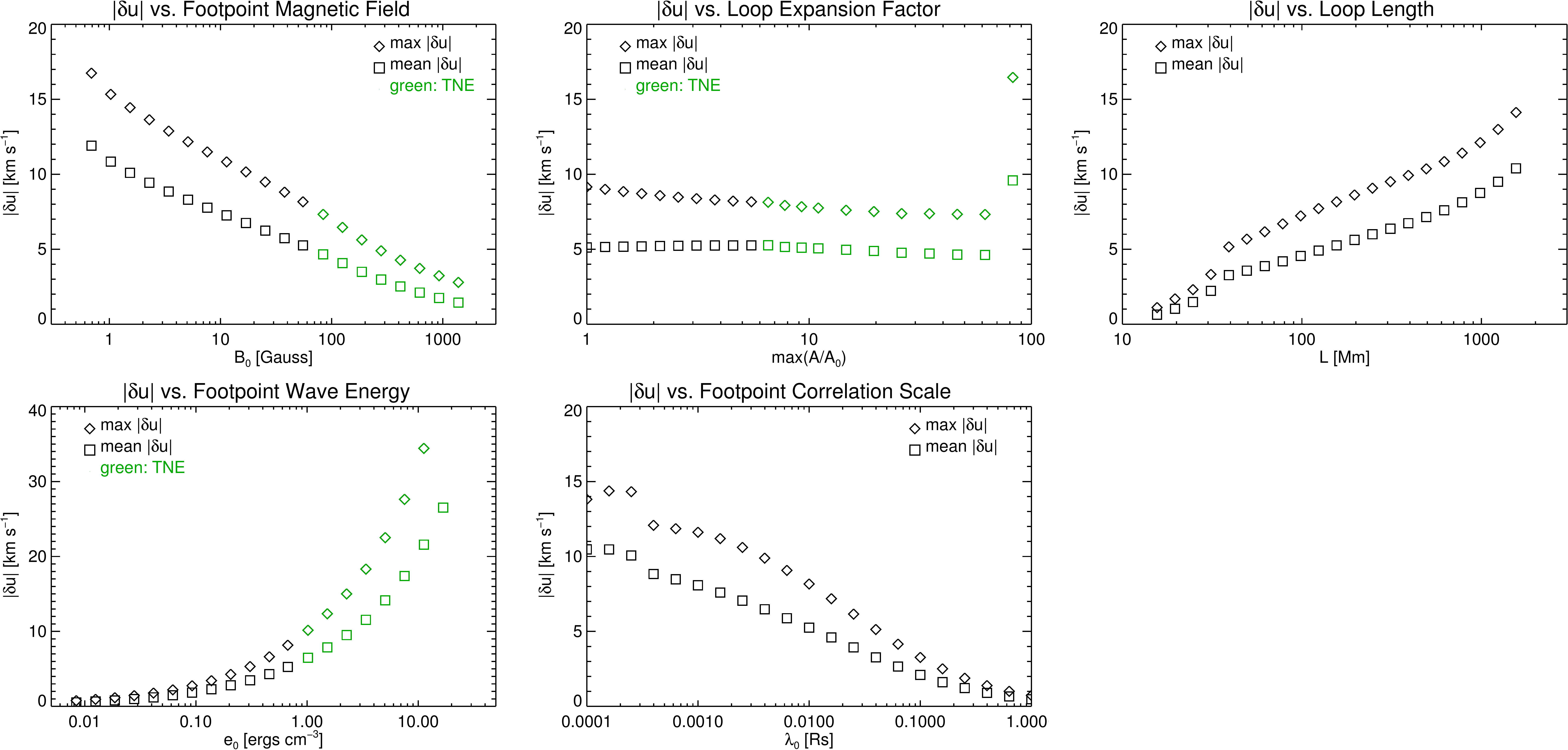}
\caption{The implied velocity fluctuation amplitude for each of the five parameter-space runs. Each panel shows the maximum value of $|\delta u|$ in diamonds, and the mean value of $|\delta u|$ along the loop in squares. The top and bottom rows show the results for the loop-dependent parameters ($B_0,\ \mathrm{max}(A/A_0),\ L$), and the independent parameters of the model ($e_0$,\ $\lambda_0$), respectively.  Note the y-axis scaling change for the bottom left panel ($e_0$).}
\label{fig:pspace_deltau} 
\end{figure*}

In \cref{fig:pspace_deltau} we show the mean of $|\delta u|$ averaged along the loop (squares), and the maximum of $|\delta u|$ (diamonds) for each of the parameter-space simulations. Apart from the case in which we vary the wave energy, where we expect $|\delta u|$ to grow roughly with $\sqrt{e_0}$, we generally see that the maximum of $|\delta u|$ stays below 15 {\kms} and the mean ranges from a few to 10 {\kms}, despite the large variation of these parameters. 

It is evident that $\delta u$ decreases as $B_0$ and $\lambda_0$ increase, which is due to the change in shape of the $z_\pm$ profiles as they transition from a regime dominated by the dissipation term to one dominated by the reflection and propagation terms. Also interesting is the increase of $|\delta u|$ with loop length.  As $L$ grows, the low coronal profiles of $z_\pm$ are set primarily by the self-reflection of the driven species, and the role of the wave launched at the other end of the loop is diminished, implying that the $z_+$ and $z_-$ profiles cancel each other less with increasing $L$, causing $|\delta u|$ to grow.

%--------------------------------------------------------------------------------------------------------------------------------------------
% Realistic Loops Study
%--------------------------------------------------------------------------------------------------------------------------------------------
\section{Scaling for Realistic Coronal Loops}
\label{section:realistic_loops}

Having explored the parameter space of the WTD heating model has been explored, we now study solutions in realistic coronal loops. In this context, `realistic' refers to loops traced from three-dimensional (3D) extrapolations of the magnetic field using observations. For such fields, the base magnetic field strengths, loop lengths, and areal expansion profiles are all set by the complex, 3D nature of the field, and there is generally no symmetry between footpoints. This complexity makes the analysis more complicated, but for a coronal heating model to be broadly applicable it must scale reasonably under realistic coronal conditions.

\subsection{Loop Properties}

\begin{figure}[hbtp]
\centering
\includegraphics[width=\cdonecol]{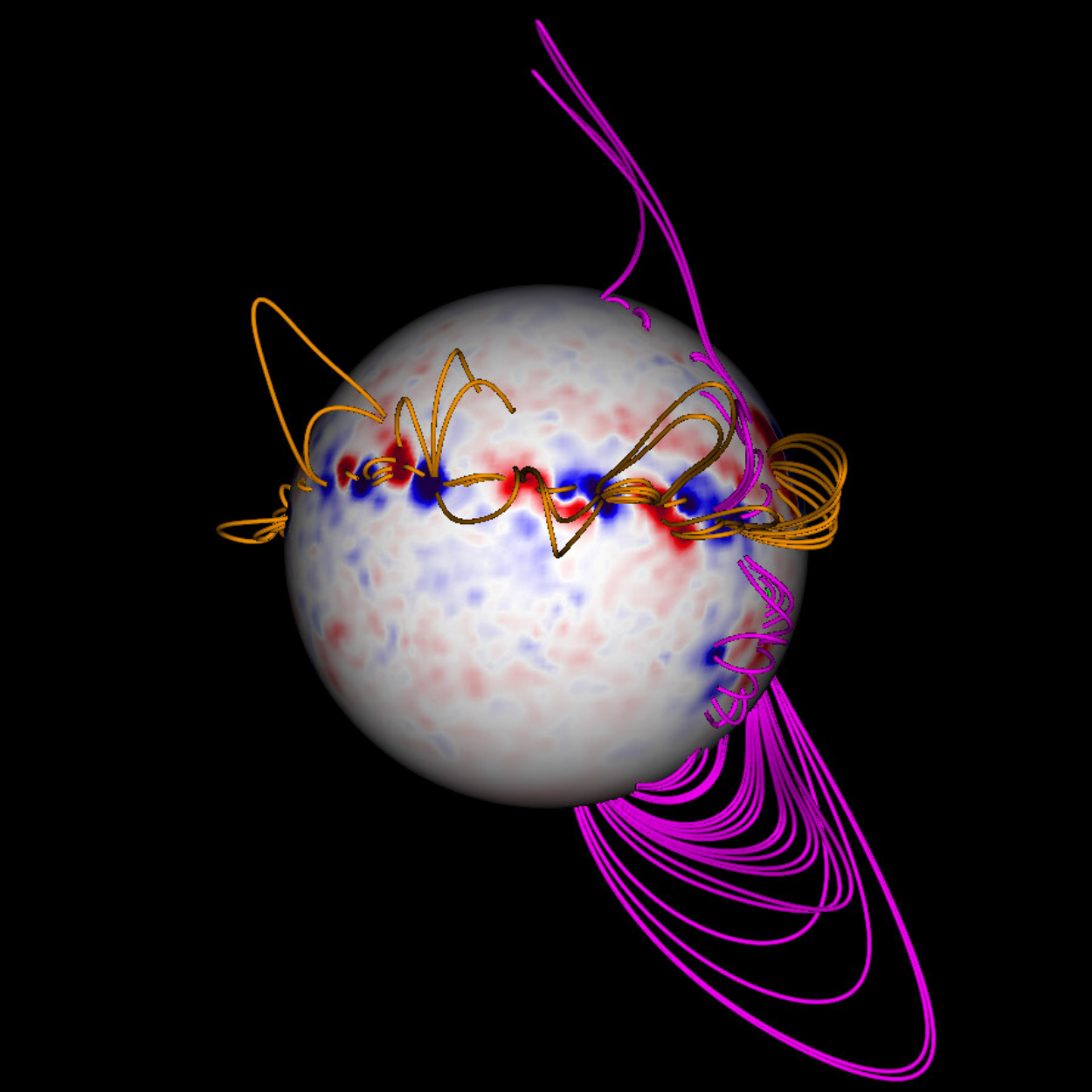}
\caption{Visualization of the `realistic' loops selected for our study as seen from Earth's perspective on 27 Nov 2011. Magenta field lines show the quiet-sun (QS) subset and gold field lines show the active region (AR) subset. The radial magnetic field distribution at the base of the corona is shown in the blue-red colormap, and is the same as described in \citet{downs13}. 
}
\label{fig:selected_field_lines} 
\end{figure}

To characterize the model over a range of quiet and active conditions we obtain field line tracings from a full-sun thermodynamic MHD model of the global corona on 2011 December 12. The details of the simulation are given in \citet{downs13}, but our familiarity with the case was the main reason for choosing it--this experiment could have easily been done using other global field models or extrapolations. The selected 98 loops are shown in Figure \ref{fig:selected_field_lines}. This subset is selected by first tracing thousands of field lines from the photosphere along a N/S arc that cuts mainly across quiet-sun (purple) regions and an E/W arc that cuts across the main belt of northern active regions at this time (gold). The traces are then sorted into bins as a function of apex height and a representative field line is chosen from each bin at random. 

\begin{figure*}[hbtp]
\centering
\includegraphics[width=\cdfullpage]{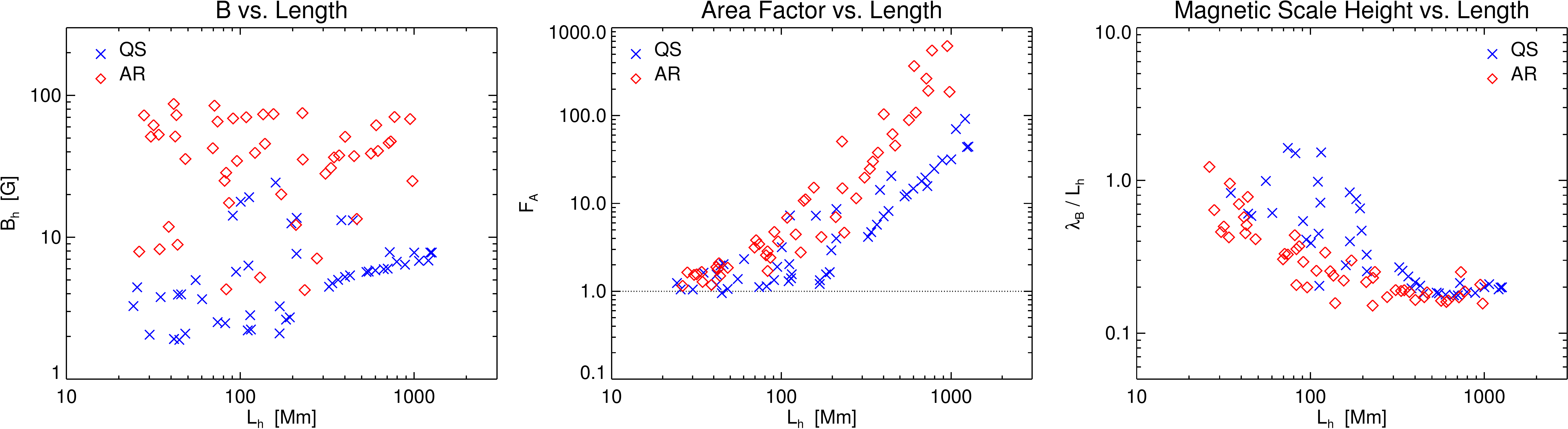}
\caption{Static properties of the selected `realistic' loops as a function of loop half-length. The quiet-sun and active region subsets are indicated with the blue crosses and red diamonds. Left: the harmonic mean of the footpoint magnetic field strengths. Middle: the area factor (Eq. \ref{eq:area_factor}), where 1.0 indicates a uniform loop. Right: the fitted scale height of the magnetic field profile along the loop divided by the loop half-length.
}
\label{fig:realistic_loops_static_properties} 
\end{figure*}

Selected parameters for the QS (blue) and AR (red) loop distributions are displayed in \cref{fig:realistic_loops_static_properties} as a function of the loop half-length, $L_h$. The left panel shows the variation in footpoint field strength, which is shown as a harmonic mean of the left and right footpoint fields, $B_h$. For our selected loops, we see that $B_h$ is generally larger for the AR loops and that both distributions are fairly flat with some random variation.

The middle panel of \cref{fig:realistic_loops_static_properties} is used to illustrate the relative expansion of the selected loops. Similar to the expression for the coronal heat flux (\cref{eq:heatflux_definition}), we can define a dimensionless area factor by integrating the area along the loop and dividing by the total area and $L_h$:
\begin{equation}
F_A = \frac{1}{L_hA_T}\int_0^L{Ads}.
\label{eq:area_factor}
\end{equation}
$F_A$ is exactly one for a uniform loop (regardless of length), and grows as the areal expansion grows\footnote{This factor is related to the `average magnetic field strength', $\left <B\right >$, defined in \citet{mandrini00}, where $\left <B\right >=B_h/F_A$.}.  In general this indicates that the realistic field lines do not have uniform cross-sections, and that the relative expansion grows with length (due to the natural decay of $B$ with height). Interestingly, we also see that the AR loops tend to have more relative expansion than their QS counterparts for a fixed $L_h$. 

In the right panel of \cref{fig:realistic_loops_static_properties} we plot the effective scale height of the magnetic field, $\lambda_B$, divided by $L_h$. This scale height is determined by fitting in the same manner as done for the heating profiles (Section~\ref{section:single_loop_caseA}), and the left and right fits are combined into a harmonic mean (displayed). We see that the relative length-scale of the magnetic field profile starts off roughly comparable to loop length for the shortest loops, but quickly falls and finally asymptotes to a factor of 1/5 or so. Because the scale height of the heating profile is intimately tied to the magnetic field profile (\cref{eq:heating_scale_analytic}), this suggests that footpoint concentrated heating may arise naturally for these loops.

\subsection{Solution Properties}

The next step is to run the WTD model on the selected loops. We set a boundary density of $n_{e,0}=$\neu{6}{12}{\un} and a driving amplitude of $z_0=10.3$ {\kms}. To ensure a sufficient run-time, each loop is run for a minimum of 40 Alfv\'en crossing times or 60 hours of physical time, whichever is longer. The time-dependent loop simulations are then processed in the same way as in Section~\ref{section:paramspace}. The majority of the loops reach a steady-state solution, but some appear to show cyclic TNE behavior.

\begin{figure}[hbtp]
\centering
\includegraphics[width=\cdonecol]{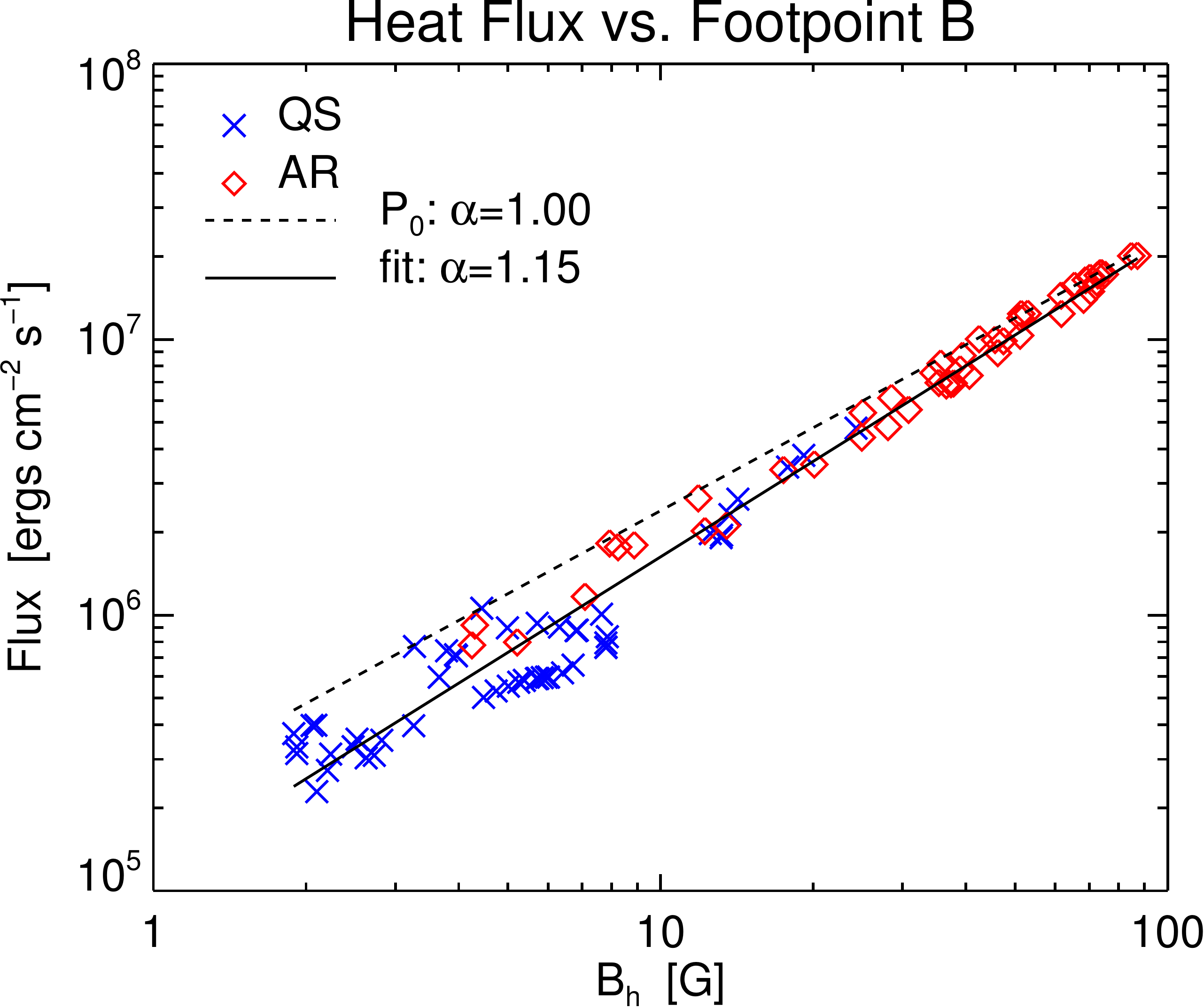}
\caption{The scaling of the heat flux deposited as a function of $b_h$ for the selection of `realistic' loops. The dotted line indicates a perfect linear scaling, while the solid line is a the best fit to the data ($\alpha=1.15$). The quiet-sun and active region subsets are indicated with the blue crosses and red diamonds.
}
\label{fig:realistic_loops_heatflux_scaling} 
\end{figure}

\cref{fig:realistic_loops_heatflux_scaling} shows the scaling of the coronal heat flux as a function of $B_h$ for the QS and AR sets of loops. The dotted line shows the input poynting flux, $P_{in}$, and the solid line is a logarithmic fit to the data ($y=x^\alpha$). The fitted scaling gives $\alpha=1.15$, which is slightly steeper than the exact scaling of $P_{in}$ with $B_h$ ($\alpha=1.0$). This is reasonable because not all of the input flux will be dissipated due to reflection and transmission (relative efficiency, $R_e\lesssim1$). In this case, it appears that the efficiency improves as $B_h$ gets larger, consistent with the scaling of $B_0$ in the parameter-space study (\cref{fig:pspace_loop_params}). This confirms the analytic expectation that the coronal heat flux scales with the magnetic field for realistic loops. This is likely a necessary requirement for a general coronal heating model \citep{fisher98,pevstov03}.

\begin{figure*}[hbtp]
\centering
\includegraphics[width=\cdfullpage]{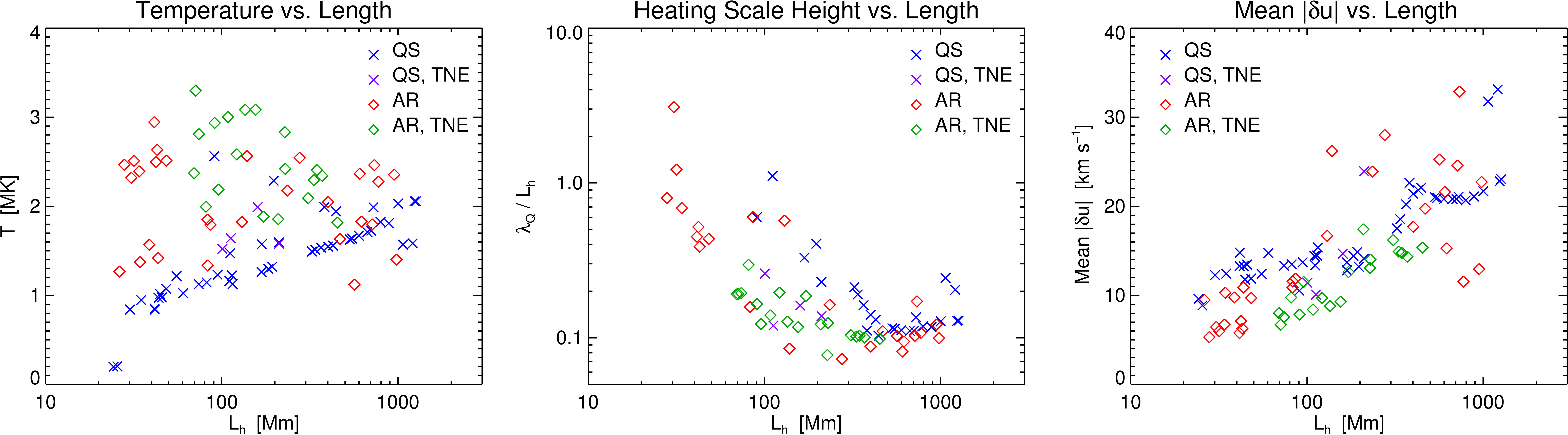}
\caption{Properties of the WTD solutions as a function of $L_h$ for the selection of `realistic' loops. Left: the maximum loop temperature.  Middle: the fitted heating scale height, $\lambda_Q$, divided by $L_h$. Right: the mean value of $|\delta u|$ averaged over the loop. The quiet-sun and active region subsets are indicated with the blue crosses and red diamonds. Loops with repeated TNE cycles are indicated in purple (QS) and green (AR). 
}
\label{fig:realistic_loops_solution_properties} 
\end{figure*}

The left panel of \cref{fig:realistic_loops_solution_properties} shows the resulting loop temperature as a function of length for both the AR and QS distributions. For solutions that reached a steady state, we plot the maximum loop temperature (blue and red), and for solutions undergoing cyclic, TNE behavior we plot the midpoint temperature of the cooling phase (purple and green). For the AR loops, we find that reasonable AR temperatures (2.5-3.5 MK) are achieved for loop half-lengths between 30 and 300 Mm. Many of these AR loops (18 of 48) undergo non-steady TNE behavior, while just a few of the QS loops do (4 of 50). The temperatures for the QS loops show an increasing trend from low to high as $L_h$ increases, and these temperatures are consistent with observations of the quiet-sun.

To assess the stratification of the heating profile, we show the ratio of the fitted heating scale height to the loop half-length in the middle panel of \cref{fig:realistic_loops_solution_properties}. These scale heights follow a similar trend to the magnetic field scale heights (albeit with more spread), confirming that the heating becomes relatively concentrated at the footpoints for $L_h\gtrsim70$Mm. Interestingly, we also find that loops with TNE solutions appear to cluster together, spanning a range of short to medium $L_h$ (70-500 Mm) with stratification ratios ($\lambda_Q/L_h$) smaller than at least 1/3. This result is consistent with the general result that hydrostatic solutions may be difficult to obtain for strongly concentrated heating \citep[e.g.][]{aschwanden01}, and this result is consistent with the idealized TNE simulations conducted by \citet{mikic13}.

Lastly, in the right panel of \cref{fig:realistic_loops_solution_properties} we show the mean of $|\delta u|$ averaged over all of the loop solutions as a function of $L_h$. Like the parameter-space exploration (Section \ref{section:paramspace:deltau}), we see an increasing trend of the mean $|\delta u|$ as a function of loop length, but now with significantly more scatter. This is to be expected because this set of realistic loops has a range of magnetic field profiles and asymmetries, changing the way in which the $z_+$ and $z_-$ profiles add along the loop. Taking the the mean and maximum values of $|\delta u|$ for each loop, and then producing a median of each quantity over all loops, gives a median value 14.0 {\kms} for the mean of $|\delta u|$ and 23.5 {\kms} for the maximum of $|\delta u|$. Although the model is too idealized to compare these numbers directly to non-thermal line widths, these results do indicate that the model velocity fluctuations are within a reasonable range, even for a broad range of realistic loop backgrounds.

\subsection{Scaling of the Local Heating Rate}

Lastly, we would like to relate the results of the WTD heating model to prior observational, theoretical and empirical work. One popular way to characterize a heating model is to phrase the heating rate (or heat flux) in the form of a scaling law as a function of some set of loop properties. For example, \citet{mandrini00}, \citet{schrijver04}, \citet{warren06}, and \citet{lundquist08} all attempted to use a combination of observations and model/simulations to determine a best-fit scaling law for the heating rate. Typically the scaling law will depend on factors such as the magnetic field strength, loop length, and density, although specific definitions, such as footpoint vs. average field strengths, may vary.  Scaling laws of this form can be compared to predictions from various theories \citep[Table 5 of][and references therein]{mandrini00}, and can be adapted for empirical use in 3D simulations that are benchmarked against observations \citep[e.g.][]{mok08,lionello09,downs10,mok16}.

In one of the more recent studies, \citet{lundquist08} looked at how the average heating rate, $\bar{Q}=\int_0^L{Q}ds/L$, scales with the average magnetic field strength, $\bar{B}=\int_0^L{|B|}ds/L$, and $L$. Testing four combinations of $\bar{Q}\propto\bar{B}^\alpha L^\beta$ with a uniform heating model, they determined that $\bar{B}/L$ scaling ($\alpha=1$, $\beta=-1$) was most consistent with their set of AR observations. They argued that their results were consistent with the constraints put forth by \citet{mandrini00} and also in line with similar work by \citet{warren06}. This result implied steeper scaling than found by \citet{schrijver04}, who argued instead that coronal heat flux should scale with $B/L$, which would imply $B/L^2$ scaling for the local heating rate for a uniformly heated, constant cross-section loop\footnote{See the discussion in \citet{lundquist08} for more details.}. 

On the surface, our results are consistent with \citet{mandrini00} and \citet{lundquist08}. We can similarly compute the average WTD heating rate, $\bar{Q}_w$, and $\bar{B}$ from our 98 selected loops. A fit to a scaling law of the form $\bar{Q}_w\propto \bar{B}^\alpha L^\beta$ gives $\alpha=1.24\pm0.24$ and $\beta=-0.59\pm0.22$, which is quite close to $\bar{B}/L$ scaling (our fitting technique is described below).  However, the explicit assumption of a uniform, time-independent heating rate in these studies makes it difficult to compare our results directly. Essentially, the assumption of uniform heating limits the available solution space to steady-state solutions without overpressure at the loop footpoints. In that sense it is worth investigating more general scaling laws where the variation of the heating rate along the loop is explicit in the formulation (i.e., depends on localized parameters of the loop).

Posing a scaling law for the volumetric heating rate of the form $Q\propto B^\alpha L^\beta$, where now $B$ is the local magnetic field strength, we can ask how the WTD-driven heating model scales for these parameters. Each loop simulation provides a range of $Q_W$ vs. $B$ values along its length and we sample every loop 400 times uniformly along $s$. Selecting all points from this group with local temperatures above our transition region broadening temperature ($T_e$ = 0.35 MK) gives a total of 34,032 values for $Q_W(B,L)$. 

\begin{figure}[hbtp]
\centering
\includegraphics[width=\cdonecol]{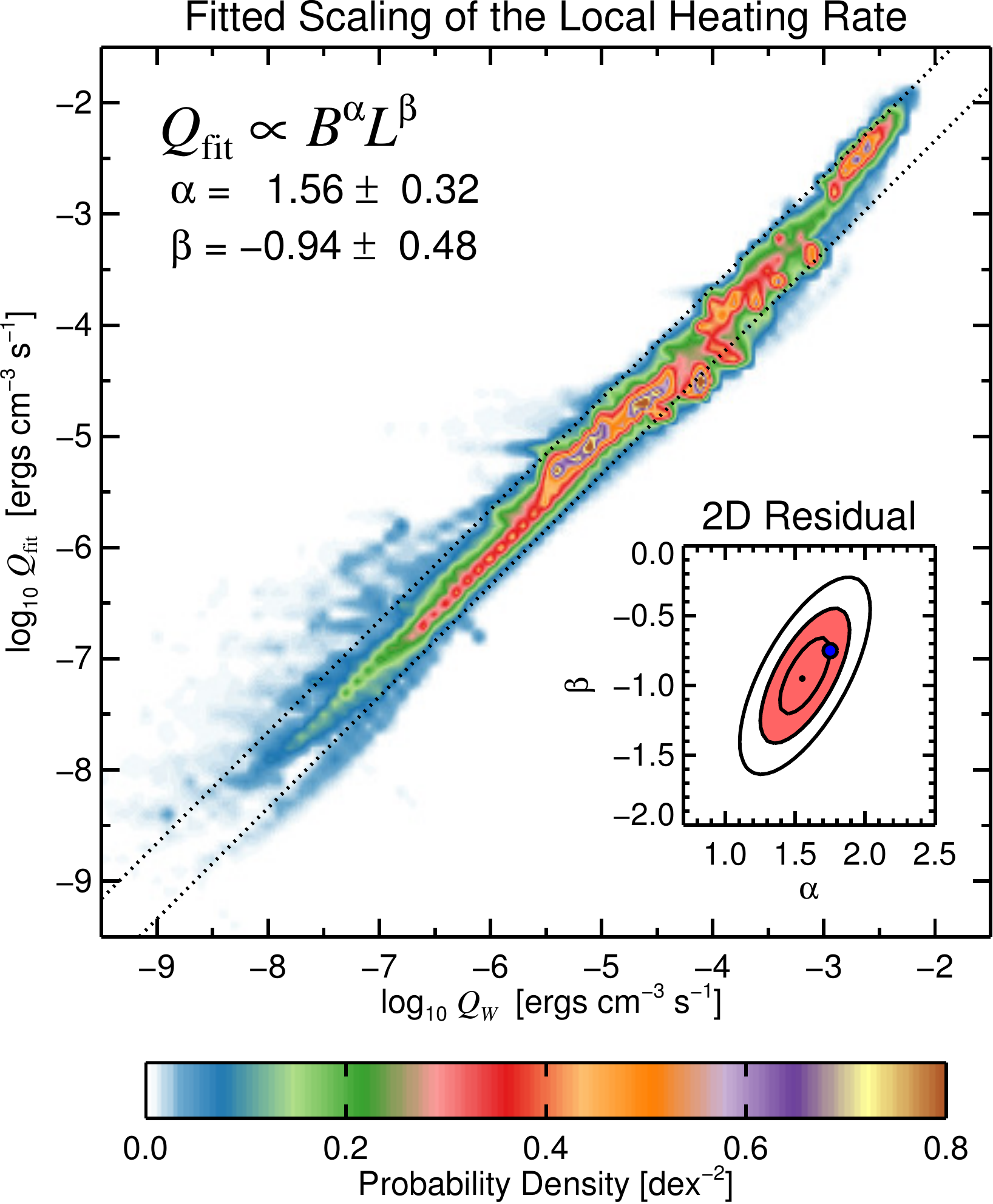}
\caption{Correlation of the local heating rate to the best fit scaling law for all 34,032 sampled points. All points are binned into a 2D histogram covering the log-log space, and the resulting probability density of this distribution is shown in color. The width of the standard deviation for the best fit ($\pm~0.30$~dex) is shown with the dotted lines. The inset shows contours of the 2D residual for every $\alpha$ and $\beta$ combination. The black dot shows the best fit minimum, and the contours are at 1.05, 1.15, and 1.30 of this minimum. The blue dot indicates the empirical scaling law from \citet{mok16}. The error bars for $\alpha$ and $\beta$ are derived from the total range of the 1.15 contour, which is shaded in red.
}
\label{fig:realistic_loops_heating_scaling} 
\end{figure}

\Cref{fig:realistic_loops_heating_scaling} shows $Q_W$ vs. the best fit scaling law, $Q_\text{fit}$, as a log-log correlation over several decades. Because of the sheer number of points, this is visualized as a color plot of the probability density derived from a 2D histogram of all points in this space. The standard deviation of $\log_{10}(Q_W/Q_\text{fit})=0.3$ dex is used to illustrate the typical width of the distribution (dotted lines). The best fit is determined through a linear regression of $\log_{10}Q_W$ as a function of $\log_{10}B$ and $\log_{10}L$, and we obtain scaling values of $\alpha=1.56\pm0.32$ for $B$ and $\beta=-0.94\pm0.48$ for $L$. The error bars shown are not determined by a standard $\chi^2$ analysis, which gives minuscule error bars because of the large number of points, but are determined instead from the 2D residual of the standard deviation in $\alpha, \beta$ space (\cref{fig:realistic_loops_heating_scaling}, inset). We take values within 1.15 of the residual minimum (red shaded region) as a metric for a visually reasonable fit, and use this to produce the error estimates.

The fitted scaling of the WTD model turns out to be quite similar to the empirical scaling law used by \citet{mok16} in a 3D simulation of realistic active region thermodynamics. They used $Q\propto B^{1.75}L^{-0.75}n_e^{0.125}$ (blue dot, \cref{fig:realistic_loops_heating_scaling}, inset), which was inspired by earlier work on the dissipation of turbulence in a magnetic braiding scenario \citep{rappazzo07,rappazzo08}. Using this scaling law, \citet{mok16} found a favorable comparison of forward modeled AR emission to extreme ultraviolet and soft X-Ray observations, and a particular emphasis was placed on the dynamic heating and cooling cycles (TNE) that result from such a stratified heating function (similar to what we find here). The TNE evolution was also studied in more detail along 1D loops extracted from the same simulation/field model by \citet{lionello13} and \citet{winebarger14}. 

The scaling consistency between $Q_W$ and the empirical scaling law is encouraging on a number of levels. First, this provides further support that heating rates determined from the WTD model are at least reasonable and can be competitive with purely empirical formulations. Second, the WTD model, by nature of its formulation (an auxiliary equation solved along the magnetic field), has fewer free parameters ($z_0$, $\lambda_\perp$) and naturally adapts the heating rate to the magnetic field conditions encountered. This property makes it much more amenable for application across a wide variety of coronal regimes (QS vs. AR) and for the heating and acceleration of the solar wind \citet{lionello14,lionello14b}.

%--------------------------------------------------------------------------------------------------------------------------------------------
% Conclusion
%--------------------------------------------------------------------------------------------------------------------------------------------
\section{Summary and Conclusion}
\label{section:conclusion}

In this paper we investigated the properties of the a wave-turbulence-driven (WTD) heating model for closed coronal flux tubes. The parametrization is simple enough to be coupled to the full hydrodynamic evolution of plasma along a 1D loop, and is applicable to multi-dimensional MHD models. Analysis of the equations and their implications for heating gave a broad characterization of the properties of the WTD model. After studying a few illustrative cases, we constructed a robust simulation framework to rapidly scan a variety of loop parameters and establish the scaling properties of the model. The model was tested on a selection of realistic quiet-sun and active region loops to determine the practical scaling of the model. We demonstrate an implicit dependence of the WTD heating rate on the geometric and plasma properties of a loop, both along its length and at its footpoints. We generally find that the model robustly describes a broad range of quiet-sun and active region conditions and compares well to empirical heating models, despite having only two free parameters. Our most relevant findings are summarized as follows: 

1:~We studied how the average heat flux (\uf) of the WTD model scales as a function of the base magnetic field strength. We determined a near linear relationship, which is consistent with prior observational studies \citep[e.g.][]{fisher98,pevstov03,lundquist08}. This property is a natural result of the heat flux being directly related to the net Poynting flux of wave energy entering the corona. A scaling law for the local volumetric heating rate was also determined ($Q\propto B^{1.56\pm0.32}L^{-0.94\pm0.48}$). This scaling compares favorably with prior empirical modeling of active regions \citep{mok16}, which was inspired by theoretical work on turbulent dissipation \citep{rappazzo08}.

2:~The scale at which the volumetric heating rate falls from the coronal base of the loop (heating scale height) naturally becomes steeper as the magnetic field strength falls and the areal expansion factor grows. It also has a weak dependence on the density scale height. Therefore, \emph{the WTD heating model adapts to the plasma properties along the loop}. This is a feature that can be difficult to capture with simple analytic or empirical prescriptions for the local heating rate. 

3:~We find that dynamic heating and cooling cycles develop naturally when steady-state solutions are not obtained with the WTD heating model. These cycles reflect thermal nonequilibrium (TNE) solutions, and result from an inherently stratified heating concentration at the loop footpoints. Unlike impulsive heating theories that are invoked to explain heating and cooling signatures \citep[e.g.][]{viall13}, the WTD heating profile remains relatively constant even during large amplitude hydrodynamic changes. This behavior is similar to the empirical heating studies of \citet{mikic13} and \citet{mok16}. 

More broadly, we believe the WTD formulation helps to fill a gap between sophisticated models of heating driven by turbulence \citep[e.g.][]{rappazzo08,van_ballegooijen11}, and strictly empirical formulations for coronal heating \citep[e.g.][]{lionello09,mok16}.  Approaches that model turbulent fluctuations directly can be challenging to generalize to large-scale MHD models of the corona, while empirical models say little about the heating mechanism itself. That the WTD formulation can produce reasonable solar wind properties \citep{verdini10,lionello14,lionello14b}, and scales well across a wide range of magnetic field conditions is also encouraging from a unification standpoint---constructing a simple, physics based model that can be applied to both open and closed field regimes for realistic 3D magnetic fields. 

By studying the behavior and scaling of the WTD model in 1D as we have done here, we are able to thoroughly characterize its heating properties and place it in context with other work. This general framework for testing and analysis could easily be applied to other heating formulations, and obviously there are other aspects that still require investigation. Going forward, we will study how active region loops heated by the WTD model are (or are not) consistent with time dependent AR emission (such as time-lags, and loop cooling times), while carefully pursuing the end-goal of multi-dimensional MHD implementations and applications \citep[e.g.][]{vanderholst14}. 

\acknowledgements
This work was supported by NASA's Heliophysics Theory Program Contract NNH12CC01C and Heliophysics Supporting Research Grant NNX16AH03G. ZM acknowledges support and participation in a team investigation on global magnetic fields at the International Space Science Institute in Bern, Switzerland.  M.V. was supported by the NASA grant NNX15AF34G "Heliospheric Origins with Solar Probe Plus---the plasma astrophysics of the solar corona and wind". The authors would like to thank the anonymous referee, whose suggestions helped improve the manuscript.

%--------------------------------------------------------------------------------------------------------------------------------------------
% Bibliography
%--------------------------------------------------------------------------------------------------------------------------------------------
\bibliographystyle{aasjournal}

\newpage
\clearpage

%--------------------------------------------------------------------------------------------------------------------------------------------
% Figures
%--------------------------------------------------------------------------------------------------------------------------------------------
\section{Figures}
\phantom{yo}

%**********************************
%*** Analytic Background ******

%****************************************************************
%*** Solutions for each type of equation                    ******

%**********************************
%*** lambda loss plot ******

%**********************************
%*** lambda loss contributions ******

%**********************************
%*** lambda loss plot for BOTH sides ******

%**********************************
%***single loop Case A******

%**********************************
%***single loop (Case B) TNE cycle******

%**********************************
%***single loop Case B******

%**********************************
%***param space, loop params *****

%**********************************
%***param space, free params *****

%**********************************
%***param space, delta u results *****

%**********************************
%*** visual display of the loops *****

%**********************************
%*** loop static properties *****

%**********************************
%*** loop heatflux scaling *****

%**********************************
%*** loop solution properties *****

%**********************************
%*** Main Heating Rate  *****

\end{document}